\documentclass[12pt]{article}
\usepackage{latexsym}
\usepackage[dvips]{color}
\usepackage{amssymb,dsfont,amsfonts}
\usepackage{graphicx}
\usepackage{epsfig}
\usepackage{amsmath}
\usepackage{mathrsfs}
 
\usepackage{xcolor}
\usepackage[arrow, matrix, curve]{xy}
\usepackage{natbib}
\usepackage[linkbordercolor=red]{hyperref}

\usepackage{hyperref}

\newtheorem{thm}{Theorem}

\newtheorem{pro}{Proposition}
\newtheorem{cor}{Corollary}
\newtheorem{lem}{Lemma}
\newtheorem{defi}{Definition}
\newtheorem{rem}{Remark}
\newtheorem{ex}{Example}

\newtheorem{tpo}{Proof of Theorem}
\newtheorem{ppo}{Proof of Proposition}
\newtheorem{cpo}{Proof of Corollary}
\newtheorem{lpo}{Proof of Lemma}

\newcommand{\esssup}{\operatorname*{\mathrm{ess\,sup}}}

\newcommand{\qed}{\mbox{ }~\hfill~$\Box$ \vspace{1ex} }

\newcommand{\rmII}{\text{\it I\kern-.08em I\,}}
\newcommand{\rmIII}{\text{\it I\kern-.08em I\kern-.08em I\,}}
\newcommand{\rmIV}{\text{\it I\kern-.08em V\,}}

\renewcommand{\epsilon}{\varepsilon}

\begin{document}

%%=============================================

%%=============================================
%\author{
%\sc Patrick Beißner
%\\
%\small Dept. of Economics\\[-2pt] \small
%\\
%Institute of Mathematical Economics\\
%Bielefeld University

%}
\title{Coherent Price Systems and Uncertainty-Neutral Valuation}
\author{Patrick Bei\ss ner\footnote{Institute of Mathematical Economics -- Bielefeld University, 33501 Bielefeld,  Germany. Email: \href{mailto:pbeissne@math.uni-bielefeld.de}{$\mathtt{pbeissne}@\mathtt{math.uni-bielefeld.de}$}. 
I thank Frank Riedel and Frederik Herzberg for collateral advice. 
Financial support ptovided by the German
Research Foundation (DFG) and the IGK
``Stochastics and Real World Models" Beijing--Bielefeld is gratefully  acknowledged.
}
}

%\vspace{+0.5cm}
\date {%{\it IMW--Working Paper} \\
%\vspace{+0.5cm}
February 27, 2012} \maketitle

\begin{abstract}\vspace{-0.2cm}
We consider fundamental questions of arbitrage pricing arising when the uncertainty  model is given by
 a set of possible mutually singular probability measures. With a  single probability model, 
 essential equivalence between the absence of arbitrage and the existence of an equivalent
 martingale measure is a  folk theorem, see \cite{hakr79}.\newline
We establish a microeconomic foundation of  sublinear price 
systems and present an extension result. In this context we introduce a prior dependent notion of marketed spaces
and viable price systems.
\newline 
We associate
 this extension with  a canonically altered concept  of equivalent symmetric martingale measure sets, in a dynamic trading framework  under absence of  prior depending arbitrage. 
We prove the existence of such sets when volatility uncertainty is modeled
 by a stochastic differential equation, driven by Peng's $G$-Brownian motions.
% Moreover, equilibria  in economies with dynamic monetary utility processes  and payoff streams are guaranteed.
\noindent
\end{abstract}
\medskip
{\footnotesize{ \textit{Key words and phrases}: mutually singular priors, uncertain volatility, sublinear expectation,  viability of sublinear price systems, arbitrage,     equivalent symmetric martingale measures set (EsMM set), symmetric martingales,  Girsanov for $G$-Brownian motion} \\
{\it \hspace*{0.6cm} JEL subject classification: G13, G14, D46, D52, C62}
}

\section{Introduction}
 In this paper we study a fundamental assumption  behind theoretical models in Finance, namely, the assumption of a known single probability measure. Instead, we allow for a   set of probability measures $\mathcal{P}$,  such that we can guarantee awareness of potential model misspecification.\footnote{The distinction between measurable and unmeasurable uncertainty drawn by \cite{kn21} serves in this paper as a starting point for modeling the uncertainty in the economy. 
 \cite{key37} later argued that single prior models cannot represent irreducible uncertainty.} We  investigate the implications of a related  and reasonable arbitrage concept. In this context we suggest a   \textit{fair} pricing principle  associated with an appropriate martingale concept. The multiple prior setting influences the price system, in terms of the simultaneous control of different null sets. This motivates a worst case pricing theory of possible means.\footnote{This was originally noted by \cite{fin33}.}
\newline
The pricing of derivatives via  arbitrage arguments plays a fundamental role in Finance. Before stating an arbitrage concept, a probability space $(\Omega,\mathcal{F},P)$ is fixed such that  marketed claims or tradeable assets with trading strategies can be defined. The implicit assumption is that the probabilities are exactly known. The Fundamental Theorem of Asset Pricing (FTAP) then asserts equivalence between the absence of $P$-arbitrage in the market model and the existence of a  consistent linear price extension so that the market model can price all contingent claims. The equivalent martingale measure is then just an alternative description of this extension via the Riesz representation theorem. 
\newline
We introduce an uncertainty model  described as a  set of possibly mutually singular probability measures (or priors). Such models are  undominated in the sense that   no probability measure exists which controls the zero sets of all other priors in $\mathcal{P}$. Our leading motivation is a general form of \textit{volatility uncertainty}. This perspective deviates from models with a term structures of volatilities, i.e.  stochastic volatility models, see \cite{hes93}. We do not formulate the volatility process of a continuous time  asset price  via another process whose law of motion is exactly known. Instead, the legitimacy of the probability law still depends on  an infinite  repetition of variable  observation, as highlighted by \cite{kol33}.
We  include this residual uncertainty by  giving  no concrete model for the stochastics of the volatility process, and instead fix a  confidence interval for the volatility variable. We refer to  \cite{av95}. 
 Very recent developments in stochastic analysis have established a complete theory in this field, a major objective of which has been the sublinear expectation operator  introduced by \cite{pe07}.
\newline
Unfortunately, a coherent valuation principle  changes dramatically when the uncertainty is enlarged by the  possibility of different probabilistic scenarios. In order to illustrate this important point,  we consider for a moment   the uncertainty  given by one probability model, i.e. $\{P\}=\mathcal{P}$. A (weak) arbitrage refers to a claim $X$ with zero cost, a $P$ almost surely positive  and with a probably strictly positive payoff.  Formally, this can be written as 
 \begin{eqnarray*}
\pi(X)\leq 0,\quad X\geq 0\: P\textnormal{-a.s.} \quad \textnormal{and } P(X>0)>0. 
 \end{eqnarray*}
The situation changes in the case of an uncertainty model described by a set of mutually singular prior $\mathcal{P}$. The second and third condition should be formulated  more carefully, because every prior $P\in \mathcal{P}$ could govern the market. We rewrite the arbitrage condition as
 \begin{eqnarray*}
\pi(X)\leq 0,\: \:\:X\geq 0\: P\textnormal{-a.s. for all } P\in \mathcal{P}\!\! \quad \textnormal{and } P(X>0)>0  \textnormal{ for some } P\in \mathcal{P}.
 \end{eqnarray*}
Accepting this new $ \mathcal{P}$-arbitrage notion one may ask for the structure of the related objects.\footnote{See Remark 3.14 in \cite{vor10} for a discussion of a weaker arbitrage definition and its implication in the $G$-framework.} Suppose we apply the same idea of linear and coherent extensions to the uncertainty model under consideration.  Coherence corresponds to strictly positive and continuous  functionals on the whole space of claims $L$ which are consistent with the given market data of marketed claims $M$. These claims can be traded frictionless and are priced by a linear functional $\pi:M\rightarrow \mathbb{R}$.
Hence, the order structure and the underlying topology of $L$ build the basis of any financial model that asks for coherent pricing. The representation of elements in the topological dual space\footnote{We discuss the precise description in the second part of Introduction and in Section 2.2. } indicates inconsistencies between positive linear price systems and the concept of $\mathcal{P}$-arbitrage. As it is usual, the easy part of establishing an FTAP is  deducing an arbitrage free market model from the existence of an equivalent martingale measure $Q\sim P\in \mathcal{P}$  an . When seeking a modified FTAP, the following question (and answer) serves to clarify the issue:
\begin{quote}
Is the existence of a measure $Q$ equivalent to some $P\in\mathcal{P}$ such that prices of all traded assets are $Q$-martingales, a sufficient condition to prevent  a $\mathcal{P}$-arbitrage opportunity?
\end{quote}
A short argument gives us a negative answer:
Let $X\in M^0\subset M$ be a marketed claim with price $\pi(X)=0$. We deduce that $E^Q[X]= 0$ since $Q$ is related to a consistent price system. Suppose  $X\in M^0\cap L_+$ with $X\geq 0$  $P$-a.s for every $P\in \mathcal{P}$ and  $X> 0$ $P'$-a.s.  for some  $P'\in \mathcal{P}$ exists. The point is now,
\begin{quote}
 with $\{P\}=\mathcal{P}$  we would observe a contradiction since $Q\sim P$ implies $E^Q[X]>0$. But $X\in M^0$ may be such that  $P'(X>0)>0$ with  $P'\in\mathcal{P}$ is mutually singular to $Q\sim P$. 
\end{quote}
This indicates that our \textit{finer} arbitrage notion is,  in general, not consistent with a linear theory of valuation. In other words, a single prior as a pricing measure is not able to contain all the information about the uncertainty.
\newline
Since our goal is to suggest a modified framework for a coherent pricing principle,  the concept of marketed claim is reformulated by a prior depending notion of possibly marketed spaces $M_P$, $P\in\mathcal{P}$. As discussed in Example 3 below, such a step is necessary to deal with  the prior dependency of the asset span $M_P$. The likeness  of marketed spaces depends on the similarity of the involved priors.   Here, the possibility of different priors creates  uncertainty for a trader who may buy and sell claims which can be achieved frictionless.   We associate a linear price system $\pi_P:M_P\rightarrow \mathbb{R}$ for each marketed space. In this context we  posit that  coherence is based on \textit{sublinear} price systems,\footnote{This price system can be seen as an envelope of the price correspondence $\pi(X)=\{\pi_P(X):X\in M_P, P\in\mathcal{P} \}$, as in \cite{cla93}.} as  illustrated this in the following example:
\newline
Let the uncertainty model consists of two priors $\mathcal{P}=\{P,P'\}$. If $P$ is the true law, the market model is given by the set of marketed claims $M_P$ priced by a linear functional  $\pi_P$. If $P'$ is the true law, we get $M_{P'}$ and $\pi_{P'}$.  For instance, constructing a claim via self-financing strategies implies an equality of portfolio holdings that must be satisfied almost surely only  for the particular probability measure. If the trader could choose between the sets $M_{P'}+M_{P}$ to create a portfolio, additivity is a natural  requirement with the consistency condition $\pi_{P'}=\pi_{P}$ on $M_{P'}\cap M_{P}$.\footnote{See \cite{he06} for a discussion.} However,  the trader is neither free to choose a mixture of claims, nor may she choose a scenario. The equality of prices  at the intersection is not intuitive, since the different priors create  a different  price structure  in each scenario. We therefore argue, that  $\sup(\pi_{P'}(X),\pi_{P}(X))$ is a reasonable price for a claim $X\in M_{P'}\cap M_{P}$ in our multiple prior framework. This  yields to  subadditivity. In contrast to the classical  law of one price, linearity of the pricing functional is merely true under a fixed prior.\footnote{Sublinearity induced by market frictions is conceptionally different. For instance, in \cite{jou99} one convex set of marketed claims is equipped with a convex pricing functional, in which case, the possibility of different scenarios is not included.}
%\footnote{ The triple $(\Omega,\mathcal{F},P)$ is a standard notion in mathematical finance, due to the axiomatization of modern probability  by \cite{kol33}. The expectation $E^P[\cdot]$ as an object at the same foundational level was considered by \cite{da18} in terms of an integration theory on  $(\Omega,\mathcal{H})$. Here, $\mathcal{H}$ is a vector lattice  of functions $X:\Omega\rightarrow \mathbb{R}$.}
\\[1em]
\textbf{Outline and results of the paper}
\newline
We begin with an economic basis for an asset pricing principle. To do so, we introduce an appropriate notion of viability and relate this to the extension of sublinear functionals. Before we give an overview of the results, we describe the primitives of the economy:
\newline
The very basic principle of uncertainty is the assumption of different possible  future states of the world $\Omega$, which is equipped with a $\sigma$-algebra $\mathcal{F}$.  In order to tackle the mutually singular priors, we need some structure in the  state space.\footnote{See \cite{bnk10b} for a discussion of different state spaces.} 
In the most abstract setting, the states of the world $\omega\in\Omega$ build a complete separable metric
 space, also known as a Polish space. The state space contains all realizable path of security prices. For the greater part of the paper, we assume $\Omega=C([0,T];\mathbb{R})$, the Banach space of continuous functions between $[0,T]$ and $\mathbb{R}$, equipped with the supremum norm.  In the most general framework, we assume a  weakly compact set of priors $\mathcal{P}$ on the Borel $\sigma$-algebra
$\mathcal{F}=\mathcal{B}(\Omega)$. This encourages us to consider the sublinear expectation operator
\begin{eqnarray*}
 \mathcal{E}^{\mathcal{P}}(X)=\sup_{P\in\mathcal{P}} E^P[X].
\end{eqnarray*}
 In our economy the Banach space of contingent claims $L^2(\mathcal{P})$ consists of all random variables with a finite variance for all
$P\in\mathcal{P}$. The primitives of our representative agent economy are given by a preference relation  in $\mathbb{A}(\mathcal{P})$, the set of convex, continuous, strictly monotone, and rational preferences on $L^2(\mathcal{P})$.
\newline
 The  topological dual space, a first candidate 
for the space of price systems, does not  consist of elements which can be represented by  a Radon-Nykodym density $Z$.  Rather, in the present framework, it may be represented by the  pairs  $(P,Z)\in \mathcal{P}\times L^2({P})$.  With this dual space we introduce our price space of special sublinear price functionals $\mathfrak{P}(\mathcal{P})$. Proposition 1 lists the  properties of such price functionals and indicates a possible axiomatic approach to the price systems inspired by coherent risk measures.
\newline
 Sublinear prices can also be motivated by the  price systems of  partial equilibria, which consists of prior depending  linear price functionals $\pi_P$ restricted to the prior depending
marketed spaces $M_P\subset L^2(P)$, $P\in\mathcal{P}$. These spaces are joined to a unified marketed space $\mathbb{M}(\Gamma)$ in terms of an orthonormal basis argument. Here, the sublinearity is already present by a consolidation operation, called $\Gamma$, which transforms the given price systems $\{\pi_P\}_{P\in\mathcal{P}}$ to \textit{one} possibly  sublinear system  extended to some coherent element in the the price space $\mathfrak{P}(\mathcal{P})$. Our scenario based viability can then model a preference free equilibrium concept in terms of consolidation of possibilities.
\newline
Our first main result, Theorem 1, gives an equivalence between our notion of scenario-based viable price systems, and the extension of sublinear functionals. Our notion of viability, which corresponds to a no trade equilibrium, is then based on sublinear prices so  that the price functional acts linearly under a local prior.
\newline
In the second part, we consider the dynamic framework on a time interval $[0,T]$ with an augmented filtration $\mathbb{F}=\{\mathcal{F}_t\}_{t\in[0,T]}$. Its special feature is its reliance
on   the initial $\sigma$-algebra, which does not contain all null sets. Assuming  \cite{nuso10}, we have that the derivative  of the quadratic variation parametrizes  the set of priors.
The implicit dynamic structure opens  the door for a theory of dynamic sublinear expectation based on  sublinear
conditional expectation operators $\{\mathcal{E}_t(\cdot)\}_{[0,T]}$.
\newline
With the sublinear conditional expectation, a martingale theory is available which represents a possibilistic model of a fair game against nature. In this fashion, the multiple prior framework  allows us to generalize the concept of equivalent martingale measures. Instead of considering \textit{one} probability measure, we suggest that the appropriate concept  is a set of priors which is related to the statistical  set of priors through a prior depending  shift $Z_P\in L^2({P})$ in the Radon-Nykodym sense:  each prior  $P\in\mathcal{P}$ is shifted by a different (state price) density. This creates a  new sublinear expectation $\mathcal{E}^\mathcal{Q}$, generated by a set of priors $\mathcal{Q}$. Furthermore we require that the asset price $S$ under  $\mathcal{E}^\mathcal{Q}$  is mean unambiguous, i.e. ${E}^Q [S_T]={E}^{Q'} [S_T]$, for all $Q,Q'\in\mathcal{Q}$. Such a property is essential for  creating a process via a conditional expectation which satisfies the classical martingale representation property, see Appendix B.3. This is  true if and only if the martingale is  \textit{symmetric}, i.e $-S$ is a  $\mathcal{E}^\mathcal{Q}$-martingale as well.  This reasoning motivates the modification of the martingale concept,  now based on the idea of a fair game under mutually singular uncertainty. The condition that the price process  $S$ is a symmetric martingale motivates qualifying the  valuation principle as  \textit{uncertainty neutral}.\footnote{In the finite state case,  \cite{dalv10} introduce  the notion of a risk adjusted set of priors.}
\newline
The principal idea of our modified notion of arbitrage, which we call $\mathcal{P}$-arbitrage, and briefly discussed at the beginning of the introduction, was introduced  by  \cite{vor10} for the $G$-expectation framework.  In Theorem 2 we  show that under no $\mathcal{P}$-arbitrage there is a one-to-one correspondence between the extensions of Theorem 1 and  (special) \textit{equivalent symmetric martingale measure} sets,
 called EsMM sets. 
We thus establish an asset pricing theory  based on a (discounted) sublinear expectation payoff.  Corollary 1 relates EsMM sets to market completeness and to different kinds of arbitrage.
\newline
Having presented Theorem 1 and Theorem 2, we continue in the same fashion as in the
classical literature with a single prior. We consider a special class of asset prices driven by $G$-Brownian motion, related to a $G$-expectation $E_G$.\footnote{
In the mathematical literature, the starting point for consideration is  a sublinear expectation space, consisting of  the  triple $(\Omega;\mathcal{H}; \mathcal{E})$, where $\mathcal{H}$ is a special space of test random variables. If the sublinear expectation space can be represented via the supremum of a set of priors, see \cite{dhp11}, one can take   $(\Omega,\mathcal{B}(\Omega),\mathcal{P})$ as
the associated \textit{uncertainty space}. The precise definition of the concept is stated in the Appendix B.} This process is a
canonical generalization of the standard Brownian motion, whereas  the quadratic variation may move almost arbitrarily in a positive interval. The related $G$-heat equation is now a fully nonlinear PDE, see \cite{pe07b}.
\newline
 We consider a Black-Scholes like market with uncertain volatilities driven by a $G$-Brownian motion $B^G$.  The stock price $S$ is modeled as a diffusion
 \begin{eqnarray*}
d S_t = \mu(t,S_t) d\langle B^G \rangle_t + \sigma(t,S_t) d B^G_t,\quad S_0=1,\quad t\in [0,T].
\end{eqnarray*}
 This  related stochastic calculus comprises a stochastic integral notion, a $G$-It\^o formula and a
 martingale representation theorem.\footnote{We apply recent results from  \cite{xsz11},  \cite{so09},  \cite{stz11} and \cite{lipe11}.} In this mutually singular prior setting,   the (more evolved) martingale representation property, related to a sublinear conditional expectation,  is not equivalent to the completeness of the model, because the volatility uncertainty is encoded in the integrator of the price process.  
For the density process we introduce an 
 exponential martingale $\{\mathtt{E}_t\}_{t\in[0,T]}$\footnote{The precise PDE description of the $G$-expectation allows to define a universal density. Note that in the more general case we have  a  prior depending family of densities.} for $G$-Brownian motion and apply a Girsanov type theorem under the $G$-expectation. We observe the following formula
 \begin{eqnarray*}
 \Psi(X)=E_G(\mathtt{E}_T X), \quad  X\in L^2(\mathcal{P}),
 \end{eqnarray*}
 where the valuation still depends  on $\Gamma$ and the interest rate is zero. Example 6 illustrates its usefulness by relating the abstract super-replication, as discussed in \cite{dema06}, to an EsMM set.
\\[1em]
\textbf{Related Literature}
\newline
We embed the present paper into the existing literature. In  \cite{hakr79}  the arbitrage pricing principle provides an economic foundation by  relating the notion of  equivalent martingale measures with a linear equilibrium price system. Risk neutral pricing, as a precursor, was discovered by \cite{cr76}. The idea of arbitrage pricing was introduced by \cite{ro76}.
\newline
 The efficient market hypothesis of  \cite{fam70} introduces  information efficiency, a concept closely related to \cite{sam65}, where the notion of a martingale reached  neo-classic economics for the first time.\footnote{ \cite{ba00} influenced the course of Samuelson's work.}
\newline
 \cite{hapl81} as well as \cite{kre81} and \cite{ya80} continued laying the foundation of arbitrage free pricing. Later, \cite{dmw90}, presented a fundamental theorem of asset pricing for finite discrete time.
In a general semimartingale framework, the notion of no free lunch with vanishing risk  \cite{desc94}
 ensured  the existence of an equivalent martingale measure in the given (continuous time) financial market.
All these considerations have in common that the uncertainty of the model is given by a single probability measure.
\newline
Moving to models with multiple probability measures,  pasting martingale measures introduces the intrinsic structure of dynamic convexity, see  \cite{ri04} and \cite{del06}.
This type of time consistency is related to recursive equations, see \cite{chep02}, which can result in nonlinear expectation and
 generates a  rational updating principle. Moreover, the  backward stochastic differential equations can model drift-uncertainty, a dynamic sublinear expectation, see \cite{pe97}. 
However in these models of uncertainty, all priors are  related to a reference  probability measure, i.e. all priors are equivalent or absolutely continuous. Moreover, drift uncertainty does not create a significant change for a valuation principle of contingent claims.
\newline
The possible insufficiency of equivalent prior models for  an imprecise  knowledge of the environment  motivates theconsideration of mutually singular priors as illustrated at the beginning of this introduction. The mathematical discussion of such frameworks can be found for instance in \cite{pe07,nuso10,bnk10}. 
 \cite{epji11} provide a discussion in economic terms. Here, the volatility uncertainty is encoded in a non-deterministic quadratic variation of the underlying noise process.
\newline
Recalling \cite{gisc89}, this axiomatization of uncertainty aversion represents a  non-linear expectation via a worst case analysis. Similarly to  risk measures, see \cite{ar99}\footnote{ \cite{mar52} postulated the importance of diversification, a fundamental principle in Finance, which corresponds to sublinearity of risk measures.},   
 the related set of  representing priors may be not equivalent to each other. This important change allows an application of financial markets  with uncertain volatility. We refer to  \cite{dema06} for a pricing 
principle of claims via a quasi sure stochastic calculus and \cite{av95} for the first intuitive considerations.
\newline
 \cite{jou95} consider a  non-linear pricing caused by bid-ask spreads and transaction costs, where the price system is extended to a linear functional. Another classical motivator for nonlinearities is related to  superhedging, see \cite{fcm07}. In \cite{acf12},  pricing rules with finitely many state  are considered.\footnote{They establish a characterization of  super-replication pricing rules via an identification of the space of frictionless claims.} A price space of sublinear functionals is discussed in \cite{alto02}. 
 We quote the following interpretation of the classical equilibrium concept with linear prices and  its meaning (see \cite{aty00}):
\begin{quote}
''A linear price system summarizes the information concerning
relative scarcities and at equilibrium approximates the possibly
non-linear primitive data of the economy."
\end{quote}
The paper is organized as follows.
Section 2 introduces the primitives of the economic model and establishes the connection between our notion of viability and extensions of price systems . Section 3 introduces the security market model  associated with the marketed space. We also  discuss the $G$-Samuelson model.
 Section 4 concludes and discusses the results of the paper and list possible extensions.
 The first part of the appendix presents the details of the model and provides the  theorem proofs. In the second part, we discuss mathematical foundations such as the space of price systems and a collection of results of  stochastic analysis and $G$-expectations.

\section{Viability and sublinear extensions of prices}
We begin by recapping the  case where uncertainty is given by  a   an arbitrary probability space $(\Omega,\mathcal{F},P)$ as it  emphasizes sensible  difference with regard to the uncertainty model posit in this paper. 
\newline
Following, we introduce the uncertainty model   as well as  the related space of contingent claims. Then we discuss the space of sublinear price functionals. The last subsection is devoted to introducing the economy, where we give an extension result, (see Theorem 1, Section 2.3).
\\[1em]
\textbf{Background: Classical viability}
\newline
 Let  there be two dates $t=0,T$, claims at $T$ are elements of the classical Hilbert lattice $ L^2({P})=L^2(\Omega,\mathcal{F},P)$. Price systems are given by linear and $L^2(P)$-continuous\footnote{The topology is induced by the $L^2(P)$-norm.} functionals. By Riesz representation theorem, elements of the related topological dual can be identified in terms of elements in $ L^2({P})$. A strictly positive functional $\Pi: L^2({P})\rightarrow \mathbb{R}$ evaluates a positive random variable $X$ with $P(X>0)>0$, such that $\Pi(X)>0$.
\newline
A  \textit{price system } consists of  a (closed) subspace  $M\subset  L^2({P})$ and
   a linear price functional $\pi:M\rightarrow \mathbb{R}$.
 The marketed space consists of  contingent claims  achievable in a frictionless manner. 
\newline
$\mathbb{A}({P})$ is the set of rational, convex, strictly monotone and $L^2(P)$-continuous
 preference relations on $\mathbb{R}\times L^2(P)$. The consistency condition for economic equilibrium is given by the concept of viability.  A price system is \textit{viable} if 
there exists a preference relation $\succsim\in \mathbb{A}({P})$ and a bundle $(\hat x,\hat X)\in \mathbb{R}\times M$ with
\begin{align*}
 (\hat x,\hat X)\in B(0,0,\pi,M)\textnormal{ and }
   (\hat x,\hat X)\succsim (x,X)\textnormal{  for all }(x,X)\in B(0,0,\pi,M),
\end{align*}
where  $B(x,X,\pi,M)=\{(y,Y)\in\mathbb{R}\times M:y+\pi(Y)\leq x+\pi(X)\}$ denotes the budget set.
\cite{hakr79} prove the following fundamental result:
\begin{quote}
 
The price system $(M,\pi)$ is viable if and only if there is an extension $\Pi$ of $\pi$ to $L^2(P)$ that  is strictly positive.
\end{quote}
Note that strict positivity implies $L^2(P)$-continuity. 
The proof is achieved by a Hahn-Banach argument and the usage of the properties of  $\succsim$  such that  $\Pi$ creates a linear utility functional and hence a preference relation in $\mathbb{A}({P}) $.

\subsection{The uncertainty model and  the space of  claims}
We begin with the underlying uncertainty model. We consider possible scenarios which  share neither the same probability measure nor the same zero sets. Therefore it is not possible to assume the existence of a given reference probability measure when the zero sets are not the same. For this reason we need some topological structure in our uncertainty model.
\newline
Let $\Omega$, the states of the world,  be a complete separable metric space equipped with a metric  
 $d:\Omega\times \Omega\rightarrow \mathbb{R}_+$, $\mathcal{B}(\Omega)$ the
Borel $\sigma$-algebra of $\Omega$ and let $\mathcal{C}_b(\Omega)$ denote the
 set of all bounded, $d$-continuous and $\mathcal{B}(\Omega)$-measurable real valued functions. 
$\mathcal{M}_1(\Omega)$ defines the set of all probability measures on
 $(\Omega, \mathcal{B}(\Omega))$.
\newline
The uncertainty of the model is given by  a weakly  compact set of probability measure
 $ \mathcal{P}\subset \mathcal{M}_1(\Omega)$.\footnote{As shown in \cite{dhp11}, the related capacity $c(\cdot)=\sup_{P\in\mathcal{P}}P(\cdot)$ is \textit{regular} if and only if the set of priors is relatively compact. Here,  regularity refers to a reasonable continuity property. In Appendix B.2, we recall some related notions. Moreover, we give a criterion for the weak compactness of $\mathcal{P}$ when it is constructed via the quadratic variation and a canonical process. } In the following example we illustrate a construction for $\mathcal{P}$, which we apply in the dynamic setting of Section 3.
\begin{ex}
We consider a time interval $[0,T]$  and  the Wiener measure $P_0$ on the state space of continuous paths starting in zero $ \Omega=\{\omega:\omega\in C([0,T];\mathbb{R}):\omega_0=0\}$ and the canonical  process $B_t(\omega)=\omega_t$. Let  $\mathbb{F}^o=\{ \mathcal{F}^o_t\}_{t\in[0,T]}$, $\mathcal{F}^o_t =\sigma(B_s,s\in[0,t])$  be  the raw  filtration
of the canonical process $B$. 
\newline
 The  strong formulation of volatility uncertainty is based upon 
 martingale laws with stochastic integrals:
\begin{align*}
P^\alpha:=P_0\circ (X^\alpha)^{-1},\quad X_t^\alpha=\int_0^t \alpha^{1/2}_s d B_s,
\end{align*}
where the integral is defined $P_0$ almost surely.  The process $\alpha$ is  $\mathbb{F}^o$-adapted and has a finite first moment. A set $\mathcal{D}$ of $\alpha$'s build $\mathcal{P}$ via the associated prior $P^{\alpha}$, such that $\{P^\alpha:\alpha\in\mathcal{D}\}=\mathcal{
P}$ is weakly compact.\footnote{In order to define universal objects, we need the pathwise construction of stochastic integrals, (see  \cite{fo81}, \cite{kar95}). }
\end{ex}
We describe the set of contingent claims. Following \cite{hust73}, for each $\mathcal{B}(\Omega)$-measurable real function $X$ such that $E^P[X]$ exists 
 for every $P\in\mathcal{P}$, we define the upper expectation operator
$\mathcal{E}^{\mathcal{P}}(X)=\sup_{P\in\mathcal{P}} E^P[X]$.\footnote{It is easily verified  that $\mathcal{C}_b(\Omega)\subset \{X\:\: \mathcal{B}(\Omega) \textnormal{-measurable}:\mathcal{E}^{\mathcal{P}}(X)<\infty\}$ holds and 
$\mathcal{E}^{\mathcal{P}}(\cdot)$ satisfies the property of a sublinear expectation. For details, see Appendix A.1.1, \cite{pe07} and Appendix B.3.} We suggest the following norm  for the space of contingent claims,  given by
the capacity norm $c_{2,\mathcal{P}}$,  defined on $\mathcal{C}^b(\Omega)$ by
\begin{eqnarray} 
c_{2,\mathcal{P}}(X)= \mathcal{E}^{\mathcal{P}}(X^2)^{\frac{1}{2}}.
\end{eqnarray} 
Define the closure of $\mathcal{C}_b(\Omega)$  under ${c_{2,\mathcal{P}}}$ norm by $\mathcal{L}^2(\mathcal{P})=\mathcal{L}^2(\Omega,\mathcal{B}(\Omega),\mathcal{P})$.\footnote{See  \cite{bnk10} for this method.} Let $L^2(\mathcal{P})=\mathcal{L}^2(\mathcal{P}) /\mathcal{N}$ be the quotient space of $\mathcal{L}^2(\mathcal{P})$ by the $c_{2,\mathcal{P}}$ null elements $\mathcal{N}$.
We do not distinguish between classes and their representatives. Two random variables  $X,Y\in L^2(\mathcal{P})$  
can be distinguished if there is a prior in $P\in\mathcal{P}$ such that $P(X\neq Y)>0$.
\newline
It is possible to define an order relation $\leq$ on $L^2(\mathcal{P})$.
Classical arguments prove that  $(L^2(\mathcal{P}),c_{2,\mathcal{P}},\leq)$ is a Banach lattice, (see Appendix A.1 for details).
\newline
We  consider the space of contingent claims $L^2(\mathcal{P})$ so that under every probability model $P\in\mathcal{P}$, we can evaluate the variance  of a contingent claim.  Properties of random variables are required to be true $\mathcal{P}$-quasi surely, i.e. $P$-a.s. for every $P\in\mathcal{P}$. This indicates that a related stochastic calculus on a probability space is unsuitable.

\subsection{Scenario-based viable price systems}
This subsection is divided into three parts. First,  we introduce a new dual space where linear and $c_{2,\mathcal{P}}$-continuous functionals are the elements. As discussed in the introduction, we allow sublinear prices as well. This forces us to extend the linear price space, where we  discuss two operations on the new price space and  take a leaf out of \cite{alto02}.\footnote{In principle there is a third operation which ignores a subset of priors.  This ignorance is in some sense redundant, since we can a priori shrink the set of priors, see Appendix B.1.1 for this  operation.} We  integrate over the set of priors for the addition operation of functionals.\footnote{This operation is associated to a weighting of priors.} In Proposition 1, we list standard properties of  coherent price functionals. The last part in this subsection focuses on the consolidation.
\\[1em]
\textbf{Linear and $c_{2,\mathcal{P}}$-continuous prices systems on $L^2(\mathcal{P})$ }
\newline
Now, we present the basis for the modified concept of viable price systems. The mutually singular uncertainty generates a different space of contingent claims.
This gives us a new topological dual space $L^2(\mathcal{P})^*$.  The discussion of the dual space is only the first step to get a reasonable notion of  viability which accounts  for the present type of uncertainty.
\newline
We introduce, the topological dual of $(L^2(\mathcal{P}),c_{2,\mathcal{P}})$.
In Appendix B.1, we give a result, which asserts that the dual space consists of special  measures:
 \begin{eqnarray*}
L^2(\mathcal{P})^*=\left\{\mu=\rho P: P\in\mathcal{P} \textnormal{ and }\rho\in L^2(P)_+  \right\}. 
\end{eqnarray*}
 This representation delivers an appropriate form of the dual space. 
The $\rho$ in the representation matches with the classical Radon-Nykodym density  of the Riesz representation when only one prior $P$ lies in $\mathcal{P}$.  The  space's description  allows for an interpretation of a state price density  based on some prior $P\in\mathcal{P}$.
The stronger capacity norm $c_{2,\mathcal{P}}(\cdot)$ in comparison to the classical single prior $L^2(P)$-norm implies a richer
dual space, controlled by the set of priors $\mathcal{P}$.
Moreover, one element in the dual space chooses implicitly a prior $P\in\mathcal{P}$ and ignores all other priors. This foreshadows the insufficiency of a linear pricing principle under the present uncertainty model, as indicated in the introduction
\\[1em]
\textbf{The price space of sublinear expectations}
\newline
In this subsection we introduce a set of sublinear functionals defined on $L^2(\mathcal{P})$.  The singular prior uncertainty of our model induces the appearance of non-linear 
price systems.\footnote{A subcone of the super order dual is considered in  \cite{alto02}.  They  introduces the mathematical lattice theoretic framework and consider the notion of a semi lattice. In \cite{aft05},  \cite{aty00} general equilibrium models with superlinear price are considered in order to discuss a non-linear theory of value. These cases relate nonlinearity in terms of personalized prices, which may be applied to a differential information economy.}
Let $k(\mathcal{P})$ be the convex closure of $\mathcal{P}$. 
We refer to this space as the \textit{coherent price space} of $ L^2(\mathcal{P})$ generated  by linear $c_{2,\mathcal{P}}$-continuous functionals:
\begin{eqnarray*}
\mathfrak{P}(\mathcal{P})\!=\!\bigg{\{}\!\Psi:\!L^2(\mathcal{P})\rightarrow\mathbb{R}:\!\Psi(\cdot)\!=\!\sup_{P\in A} E^P[Z_P\cdot] \textnormal{ with }A\subset k(\mathcal{P}), Z_P\in L^2(P)_+  \bigg{\}}
\end{eqnarray*}
Elements in $\mathfrak{P}(\mathcal{P})$ are constructed by a set  of $c_{2,\mathcal{P}}$-continuous linear functionals $\{\Pi_P:L^2(\mathcal{P})\rightarrow\mathbb{R}\}_{P\in \mathcal{P}}$, 
 which are consolidated by a point-wise maximum.  We illustrate this in the following example, for details see Appendix A.1.1.
\begin{ex}
Let $\{A_n\}_{n\in\mathbb{N}}$ be a  partition of $\mathcal{P}$. And let $\mu_n$ be a positive measure on $A_n$
with $\mu_n(A_n)=1$. The resulting prior $P_n(\cdot)=\int_{A_n}P(\cdot)d\mu_n $ is given  by a weighting  operation $ \Gamma_{\mu_n}$. When we apply $\Gamma$ to the density we get $Z_n=\int_{A_n}Z_P d\mu_n $. Then, these new prior density pairs $(Z_n,P_n)$ can be consolidated by  the supremum operation of the expectations $E^{P_n}[Z_n\cdot]$. 
\end{ex}
 A full lattice theoretical discussion of our price space $\mathfrak{P}(\mathcal{P})$ lies beyond the scope of this paper.  
 The following proposition discusses properties and the extreme case of functionals in the price space $\mathfrak{P}(\mathcal{P})$.
\begin{pro}
Every functional in $\mathfrak{P}(\mathcal{P})$ satisfies the following properties:
\begin{enumerate}
\item \textit{Sub-additivity}: $\Psi(X+Y)\leq \Psi(X)+\Psi(Y)$ for all $X,Y\in  L^2(\mathcal{P})$
\item \textit{Positive homogeneity}: $\Psi(\lambda X)=\lambda\Psi(X)$  for all $ \lambda\geq 0$, $X\in  L^2(\mathcal{P})$
\item \textit{Monotonicity}: If $X\geq Y$ then $\Psi(X)\geq\Psi(Y)$ for all $X,Y\in  L^2(\mathcal{P})$
\item \textit{Constant preserving}:
$\Psi(c) = c$ for all $c\in \mathbb{R}$
\item \textit{ $c_{2,\mathcal{P}}$-continuity}: Let $(X_n)_{n\in\mathbb{N}} $ converge in $c_{2,\mathcal{P}}$ to some $X$, then we have $\lim_n \Psi(X_n)= \Psi(X) $.
\end{enumerate}
Moreover,  for every $P\in\mathcal{P}$ and positive measure $\mu$ with $\mu(\mathcal{P})\leq 1$, we have the following inequalities for every $X \in L^2(\mathcal{P})$
\begin{eqnarray*}
E^{P}[Z X]\leq \sup_{P'\in\mathcal{P}} E^{P'}[Z_{P'} X]\geq E^{P_\mu}[Z_\mu X],\: where\: P_\mu(\cdot)=\int_{\mathcal{P}} P(\cdot) d\mu(P).
\end{eqnarray*}

\end{pro}
Below, we introduce the consolidation operation $\Gamma$ for the prior depending price systems.  $\Gamma(\mathcal{P})$ refers to  the set of priors in $\mathcal{P}$ which are relevant. In Example 2, we  observe $\Gamma(\mathcal{P})=\mathcal{P}$.
\\[1em]
\textbf{Marketed spaces and scenario-based price systems }
\newline
In the spirit of \cite{aft05} our commodity-price duality is given by the following pairing $ \langle (L^2(\mathcal{P}),c_{2,\mathcal{P}}),\mathfrak{P}(\mathcal{P})\rangle$.
\newline
For the single prior framework,   viability  and the extension of the price system are associated with each other. This structure allows  only  linear prices and corresponds in our framework  to  consolidation  via the Dirac measure $\delta_{\{P\}}$ for some $P\in\mathcal{P}$. In this case we have $\Gamma(\mathcal{P})=\{P\}$.
\newline
We begin by  introducting  the  marketed  $L^2(P)$-closed subspaces $M_P \subset L^2({P})$, $P\in\mathcal{P}$. The underlying idea  is that  any claim in $M_P $ can be achieved, whenever $P\in\mathcal{P}$ is the true  probability measure.
This input data  resembles a partial equilibrium, depending on the prior under consideration.\footnote{One may think that a countable set of scenarios could be sufficient. As we mention in Appendix B 2, the norm can be represented via different  countable dense subsets of priors. However, for the marketed space we have a direct prior dependency of all elements in $\mathcal{P}$. This implies that different choices of countable and dense scenarios can deliver different price systems, see Definition 1 below.} Claims in the marketed space $M_P$ can be bought and sold, whenever the related prior governs the economy.  We illustrate this in the following example.
\begin{ex} Suppose the set of priors is constructed by the procedure in Example 1.
Let the marketed space be generated by the quadratic variation    of    representing an uncertain asset with payoff, at time $T$, $\langle B \rangle_T$ and  a  riskless asset $1$.  We have by construction $\langle B\rangle_T=\int_0^T \alpha_s ds$  ${P^\alpha}$-a.s., the marketed space under ${P^\alpha}$  given by 
\begin{eqnarray*}
M_{P^\alpha}=\bigg{\{}X\in L^2(P^\alpha):X= a\cdot \int_0^T \alpha_s ds +b\cdot 1 \: {P^\alpha}\textnormal{-a.s.},\: a,b\in \mathbb{R} \bigg{\}}.
\end{eqnarray*}
But $ \langle B \rangle$ coincides with the $P$-quadratic variation under every martingale law  $P\in\mathcal{P}$ $ {P} $-a.s. Therefore a different $\hat \alpha$ builds a different marketed space $M_{P^\alpha}$. Suppose $\alpha= \hat \alpha $ $P_0$-a.s. on $[0,s]$  for some $s\in]0,T]$ then we have $M_{P^\alpha}\cap M_{P^{\hat \alpha}}$ consists also of non trivial claims. Note, that $P^{ \alpha}$ and $P^{\hat \alpha}$ are neither equivalent nor mutually singular.\footnote{The event $\{\omega:\langle B\rangle_r(\omega)=\int_0^r\alpha_t(\omega) dt, r\in[0,s]\}$ has for both priors the same  positive mass but the priors restricted to the complement are mutually singular. We refer to Example 3.7 in \cite{epji11} for a similar example.}
\end{ex}
We fix linear functionals $\pi_P$ on $M_P$.  It is possible that the two components  $\pi_{P_1},\pi_{P_2}\in\{\pi_P\}_{P\in \mathcal{P}}$  have a common domain, i.e $M_{P_1} \cap M_{P_2}\neq \emptyset$. In this case it is possible to observe different evaluations between different priors, i.e $\pi_{P_1}(X)\neq \pi_{P_2}(X)$ with $X\in M_{P_1} \cap M_{P_2}$. Moreover, the set  $\{\pi_P\}_{P\in \mathcal{P}}$ of linear scenario-based price functionals inherit the uncertainty
 of the model. In the single prior setting incompleteness means $M_P\neq L^2(P)$. Note that $\Omega$ is separable by assumption, hence  $L^2(P)=L^2(\Omega,\mathcal{B}(\Omega),P)$ is a separable Hilbert space\footnote{In terms of  Example 2, $P_0$ is the  Wiener measure. In this situation, $L^2(P_0)$ can be  decomposed via the Wiener chaos expansion. The same can be done  for the canonical process $X^\alpha$ related to some $P^\alpha$. So we  can generate an orthonormal basis  for each $L^2(P^{\alpha})$, with $\alpha\in\mathcal{D}$. } for each $P\in\mathcal{P}$ and  admits a countable orthonormal basis. $M_P\otimes  M_{P'}$ refers to the linear hull of the involved  basis elements in  $M_P$ and $M_{P'}$. 
\begin{defi} 
Fix $L^2(P)$-closed subspaces $\{M_P\}_{P\in \mathcal{P}}$ with $M_P\subset L^2(P)$ and
 a set $\{\pi_P\}_{P\in \mathcal{P}}$ of linear scenario-based price functionals $\pi_P:M_P\rightarrow \mathbb{R}$.
Let the $\Gamma(\mathcal{P})$-\textit{marketed space} be given by $c_{2,\mathcal{P}}$-closure of $\Gamma$-relevant pasted marketed spaces 
\begin{align*}
\mathbb{M}(\Gamma)= \overline{\otimes_{P\in \Gamma(\mathcal{P})} M_P \cap L^2(\mathcal{P})}^{c_{2,\mathcal{P}}}.
\end{align*}
 A \textnormal{price system} for $(\{M_P,\pi_P\}_{P\in\mathcal{P}},\Gamma)$ is
 a functional  $\psi: \mathbb{M}(\Gamma)\rightarrow \mathbb{R}$, where  the  consolidation operator $\Gamma$ maps $\{\pi_P:M_P\rightarrow \mathbb{R}\}_{P\in\mathcal{P}}$ to $\psi$.

\end{defi}
The $\Gamma(\mathcal{P})$-\textit{marketed space} refers to the  space of all \textit{possible} marketed claims in the domain of the consolidation operator $\Gamma$, which is a mixture of convex combination and pointwise supremum.
For each $P$, the related marketed space $M_P$ consist of  contingent claims which can be achieved frictionless, when $P$ is the true law.   We have a set of different 
price systems $ \{\pi_P:M_P\rightarrow\mathbb{R}\}_{P\in \mathcal{P}}$.
 If we want to establish a consolidation of the scenarios in a normative
sense we need an additional ingredient in the market, namely $\Gamma$. This consolidation determines  the set of relevant priors and  therefore influences the whole marketed space.

\subsection{Preferences and the economy }
Having discussed the commodity price dual and the role of the consolidation of linear price systems, we introduce agents which are characterized  by their preference of trades on $\mathbb{R}\times L^2(\mathcal{P})$. There is a single consumption good, a numeraire, which  agents will consume at $t=0,T$. Thus, bundles $(x,X)$ are elements in $\mathbb{R}\times L^2(\mathcal{P})$, which are the units at time zero and time $T$ with uncertain outcome.
We call the set of rational preference relations $\succsim$ on $\mathbb{R}\times L^2(\mathcal{P})$, $\mathbb{A}(\mathcal{P})$, which satisfy convexity, strict monotonicity,  and $c_{2,\mathcal{P}}$-continuity.\footnote{ The class of variational preferences, axiomatized in \cite{mmr06}, may represent such preferences under mild assumption on the utility index $u$ and the penalty functional $c:\mathcal{P}\rightarrow [0,\infty]$. For instance, the domain of $c$ must be a subset of $\mathcal{P}$.} Let 
\begin{align*}
 B(x,X,\psi, \mathbb{M}(\Gamma))=\{ (y,Y)\in \mathbb{R}\times \mathbb{M}(\Gamma) :y+ \psi(Y)\leq x+ \psi(X)  \}.
\end{align*}
denote  the \textit{budget set} for a price functional 
$\psi:\mathbb{M}(\Gamma) \rightarrow\mathbb{R}$.  We are ready to define the appropriate notion of viability.
\begin{defi}
 A price system is \textnormal{scenario based viable} if 
there exists  a preference relation $\succsim\in \mathbb{A}(\mathcal{P})$ and a consumption bundle $(\hat x,\hat X)\in \mathbb{R}\times  \mathbb{M}(\Gamma)$ with
\begin{eqnarray*}
 &&(\hat x,\hat X)\in B(0,0,\psi, \mathbb{M}(\Gamma)),\\
 and && (\hat x,\hat X)\succsim (x,X), \textit{ for all }(x,X)\in B(0,0,\psi,\mathbb{M}(\Gamma)).
\end{eqnarray*}
\end{defi}
The conditions are necessary and sufficient as a classical model for an economic equilibrium, when we  find a preference relation.
Now, we relate the viability of $ (\{M_P,\pi_P\}_{P\in \mathcal{P}},\Gamma)$ with  price functionals in $\mathfrak{P}(\mathcal{P})$   defined on  the whole space $L^2(\mathcal{P})$. 
We introduce the notion of strictly positive functionals in  $\mathfrak{P}(\mathcal{P})$, namely  $\mathfrak{P}(\mathcal{P})_{++}$.  Such a functional $\Psi:L^2(\mathcal{P})\rightarrow  \mathbb{R}$ is called \textit{strictly positive} if we have $\Psi(X)>0$ for every $X\in L^2(\mathcal{P})_+$ with $P(X>0)>0$  for some $P\in\mathcal{P}$.
\begin{thm}
 A price system  $(\{M_P,\pi_P\}_{P\in \mathcal{P}},\Gamma)$ is scenario based viable if and only if there is an extension of $\psi: \mathbb{M}(\Gamma)\rightarrow \mathbb{R}$ to all of $L^2(\mathcal{P})$  in $\mathfrak{P}(\mathcal{P})_{++}$.
\end{thm}
This characterization of scenario-based viability takes scenario-based marketed spaces $\{M_P\}_{P\in \mathcal{P}}$ as given. Moreover, the consolidation operator $\Gamma$ is a given characteristic of the market. With this in mind one should think that in a general equilibrium system the locally given prices $\{\pi_P\}_{P\in \mathcal{P}}$ should be part of it. The extension we  perceived can be seen as a regulated and coherent price system for every claim in $L^2(\mathcal{P})$.
\newline
The proof of the theorem is based on the nonlinear separation of the  convex "better off" set,  and the  budget set. In principle, $\Gamma$ builds a convex functional such that one of the convex sets lies in the epigraph of $\Psi$ and the other does  not.\footnote{One may ask if a  separation  prior by prior is possible as well. This is in general not possible. We illustrate this as follows. A  prior depending pricing implies that $\psi$ must be restricted to $M_P$, for each $P\in\mathcal{P}$, which we call $\psi_P$. $\pi_P$ as a pricing is not sufficient since, $M_P \cap M_{\hat P}\neq \emptyset$ is possible for some $P\neq \hat P$, see footnote 22. But an extension of $\psi_P$ to all of $L^2(\mathcal{P})$ may now depends on the prior. } 
\newline
 In comparison to  the single prior case, the degree of incompleteness depends on the  prior under consideration.\footnote{This can be seen as an uncertainty in the given partial equilibrium.} As described in Example 2, this is a natural situation. As such, prior depending prices $\pi_P$ are also plausible. The expected payoff as a pricing principle depends on the prior under consideration, as well. This concept of scenario-based prices accounts for every $\Gamma$-relevant price system simultaneously. 
We have two operations which constitute the distillation of uncertainty.
This consolidation is a characterization of the Walrasian auctioneer, in which case  diversification should be encouraged.
 But this is the  sublinearity property.
\begin{rem}
One may ask which $\Gamma$ is appropriate. Such a question is related to the concept of mechanism design. The market planner can choose a consolidation, which influences the indirect utility of a reported preference relation. However, the full discussion of issues  lies beyond  the scope of this paper.
\end{rem}

\section{Security markets and $\mathcal{E}$-martingales}
We extend the primitives with trading dates and trading strategies. We consider a time interval where the market consists  of a riskless  security and a security with uncertain volatility  leading to the set of mutually singular priors.
We then introduce a financial market consistent with the volatility, and discuss the modified notion of arbitrage and the equivalent martingale measure. In Section 3.3, Theorem 2  associates scenario-based viability with EsMM sets.
\newline
In the last section we consider our model in the $G$-framework.  Here, the uncertain security process is driven by a $G$-It\^o process, which  shows that the concept of martingale measure sets is not an empty one.
\\[1em]
\textbf{Background: risk neutral  asset pricing with one prior}
\newline
In order to introduce  dynamics and trading dates, we fix a  time interval $[0,T]$ and a filtration $ \mathbb{F}=\{\mathcal{F}_t\}_{t\in[0,T]}$  on  $(\Omega,\mathcal{F},P)$.  Fix  an $ \mathbb{F}$-adapted asset price $\{S_t\}_{t\in[0,T]}=S\in L^2(P\otimes dt)$ and a riskless bond $S^0\equiv 1$. We next review some  terminology. 
\newline
The portfolio process of a strategy $\eta$ is called $X(\eta)$. Simple  self-financing  strategies are  piecewise constant  $ \mathbb{F}$-adapted  processes $\eta$ such that $dX(\eta)=\eta dS$, which we call $\mathcal{A}(P)$. 
A $P$-arbitrage in $\mathcal{A}(P)$   is a strategy (with zero initial capital) such that $X(\eta)_T \geq0$ and $P(X(\eta)_T>0)>0$.
\newline 
A claim is  marketed, i.e. $X\in M$, if there is a   $\eta\in \mathcal{A}(P)$ such that $X=\eta_T S_T$, then  we have the (law of one) price $\pi(X)=\eta_0 S_0$.
An equivalent martingale measure $Q$ must satisfy that  $S$ is  a $Q$-martingale and  $Q=\rho P$, where   $\rho\in L^2(P)$ is a Radon Nykodym-Density with respect to $P$. Theorem 2 of \cite{hakr79} states the following
\begin{quote}
 Under no $P$-arbitrage, there is a one to one correspondence between the continuous linear and strictly positive extension of $\pi:M\rightarrow \mathbb{R}$ to $L^2(P)$ and the equivalent martingale measure. The relation is  given by
 $Q(B)=\Pi(1_B)$ and $\Pi(X)=E^{Q}[X]$, where $B\in\mathcal{F}_T$ and $X\in L^2(P)$.
\end{quote}
This result can be seen as a preliminary version of the first fundamental theorem of asset pricing.

\subsection{The financial market with uncertain volatility}
 We specify the mathematical framework and the modified notions, such as arbitrage.
Our probability model is related to the existence of a canonical process with a modified absolutely continuous quadratic variation. We begin by modeling the market and considering the concrete construction for the set of priors. Following,  we reviewl the martingale notion for conditional sublinear expectation.  
\subsubsection{The dynamics and martingales under sublinear expectation}
The principle idea is to  transfer the result of Section 2 into a dynamic setup. The specification  in Example  1 of Section 2.1 serves as our uncertainty model.
 One can directly observe in which sense the quadratic variation creates  uncertain volatility from the construction.  We introduce the sublinear expectation $\mathcal{E}:L^2(\mathcal{P})\rightarrow \mathbb{R}$ given by the supremum of expectations of $\mathcal{P}=\{P^\alpha: \alpha\in\mathcal{D} \}$. Moreover, we assume that the set of priors is stable under pasting. For details, we refer to Appendix A.2.1.
\newline
 As we aim to equip the financial market with a dynamic structure of conditional sub-linear expectation, we introduce the information structure of the financial market as given by an augmented  filtration $\mathbb{F}=\{ \mathcal{F}_t\}_{t\in[0,T]}$, (see Appendix A.2.1 for details).
 The setting is based on  dynamic sublinear expectation terminology as instantiated by \cite{nuso10}.
\newline
We give a generalization of Peng's $ G$-expectation as an  example, satisfying the weak compactness of $\mathcal{P}$ when the sublinear expectation is represented in terms of a supremum of linear expectations.   In  Section 3.3 and in  Appendix B.3, we  consider the normal $G$-expectation  in more details. That said, a possible association of results in Section 2 depends heavily on the weak compactness of the generated set of priors $\mathcal{P}$.
\begin{ex}
Suppose a trader is confronted with a pool of models describing  volatility, as described in  \cite{hes93}. After a statistical analysis of the data two models remain plausible $P^\alpha$ and $P^{ \hat \alpha}$. Nevertheless, the implications for the trading decision deviate considerably. Even the asset span on its own depends on each scenario, (see Example 3). A mixture of both models does not change this uncertain situation at all. In order to deal with the possibilistic issue let us define the universal extreme cases  $\underline\sigma_t=\inf(\alpha_t, \hat \alpha_t)$ and $\overline\sigma_t=\inf(\alpha_t, \hat \alpha_t)$.
When thinking about  reasonable \textnormal{uncertainty management}, no scenario should be ignored.  The uncertainty model which accounts for all  cases between  $\underline\sigma$ and  $\overline\sigma$   is given by
\begin{align*}
\mathcal{P}=\{ P^\alpha: \alpha_t\in [\underline\sigma_t, \overline\sigma_t]  \textit{ for every }t\in [0,T] P_0\otimes dt\: a.e.\}.
\end{align*}
The  construction of a sublinear conditional  expectation is achieved in \cite{nu10}. Here the deterministic bounds of the $G$-expectation are replaced by path dependent bounds.\footnote{
This framework is in principle included in \cite{epji11}.  In this setting drift and volatility uncertainty are considered simultaneously.
Drift uncertainty or $\kappa$-ambiguity are well known terms in financial economics. A coherent theory, known as $g$-expectation,  is available under Brownian information. \newline
The approach is formulated via a correspondence which controls the feasible of Girsanov kernels and the derivative of the quadratic variation at once. For a model concerning drift uncertainty we refer to Section 2.2 of \cite{chep02}.}
\end{ex}
We introduce an appropriate concept for the dynamics  of the continuous time multiple prior uncertainty model. The associated objectives are trading dates, the information structure and  the price process (as the carrier of the uncertainty). In order to introduce the price process $S=\{S_t\}_{t\in[0,T]}$ of an uncertain and long lived security, we must introduce  further primitives. Define the time depending set of priors
\begin{align*}
\mathcal{P}(t,P)^o=\{ P'\in\mathcal{P}:P=P'\textnormal{ on } \mathcal{F}^o_t\}.
\end{align*}
This set of priors consists of all extensions $P:\mathcal{F}^o_t\rightarrow [0,1]$ from $\mathcal{F}^o_t$ to $\mathcal{F}=\mathcal{B}(\Omega)$ in $\mathcal{P}$. This is the set of all probability measures in $\mathcal{P}$ defined on $\mathcal{F}$ that agree with $P$ in the events up to time $t$.
Fix a contingent claim $X\in L^2(\mathcal{P})$.
In \cite{nuso10}, the unique existence of a sublinear expectation  $\{\mathcal{E}(X)_t\}_{t\in[0,T]}$ is proved by the following construction\footnote{
Representations of such martingales can be formulated via a 2BSDE.
This concept is introduced for example in \cite{stz10}.}:
\begin{align*}
\mathcal{E}(X)^o_t=\esssup_{Q'\in\mathcal{P}(t,P)^o}E^{Q'}[X|\mathcal{F}_t]\quad
\textit{P-a.s.},\quad \lim_{r\downarrow t}\mathcal{E}(X)^o_r=\mathcal{E}(X)_t.
\end{align*}
With the sublinearity of  the dynamic
sublinear conditional expectation we can define a martingale similarly to the  single prior setting.\footnote{For the multiple prior case with equivalent priors we refer to \cite{ri09}.}
\newline
The nonlinearity implies that if a process $X=\{X\}_{t\in[0,T]}$ is a martingale under $\{\mathcal{E}(\cdot)_t\}_{t\in[0,T]}$ then $-X$
is not necessarily a martingale. Despite this being the case, we call the process a \textit{symmetric} martingale. In the next subsection
we discuss their relationship to asset prices under a modified sublinear expectation.

\subsubsection{The primitives of the financial market and arbitrage}
For the sake of simplicity, we assume that the riskless asset is $S^0_t=1$, for every $t\in[0,T]$, i.e. the interest rate is zero.
 We call the related abstract financial market $\mathcal{M}(1,S)$  on the filtered space uncertainty space
$(\Omega,\mathcal{F},\mathcal{P}; \mathbb{F})$, whenever the process $S=\{S_t\}_{t\in[0,T]}$ satisfies
\begin{align*}
 S_t\in L^{2}(\mathcal{P}) \textnormal{ for every } t\in[0,T] \textnormal{ and }    \mathbb{F}\textnormal{-adaptedness}.
\end{align*}
A \textit{simple trading strategy}\footnote{As mentioned  in \cite{hapl81}  simple strategies rule out the introduction of doubling strategies and hence  the notion of admissibility.} is a $\mathbb{F}^o$-adapted stochastic process $\{\eta_t\}_{t\in[0,T]}$ in  $L^2(\mathcal{P})$ when 
there is a finite sequence of dates $0<t_0\leq\cdots \leq t_N=T$ such that $\eta=(\eta^{(0)},\eta^{(1)})$ can be written with  $ \eta^i \in L^2(\Omega,\mathcal{F}_{t_i},\mathcal{P})$ as $\eta_t=\sum_{i=0}^{N-1} 1_{[t_{i+1},t_{i}[} (t)\eta^i$.
\newline
The fraction invested in the riskless asset is denoted by $\eta_t^{(0)}$, $t\in[0,T]$.
A trading strategy is \textit{self-financing} if $\eta^{(0)}_{t_{n-1}}S^0_{t_n}+\eta^{(1)}_{t_{n-1}}S_{t_n}=\eta^{(0)}_{t_{n}}S^0_{t_n}+\eta^{(1)}_{t_{n}}S_{t_n}$ for every
 $n=1,\ldots,N$. The value of the portfolio  $X(\eta)$ takes values in $ L^2(\mathcal{P})$ for every  $t\in[0,T]$.
\newline 
The set of simple self-financing trading strategies is denoted by $\mathcal{A}$.	
 This financial market  $\mathcal{M}(1,S)$ with trading strategies in $\mathcal{A}$ is called $\mathcal{M}(1,S,\mathcal{A})$. 
\newline
It is well known that a necessary condition for  equilibrium is the absence of arbitrage.
 Therefore, with regard to the equilibrium result of the last section, we introduce arbitrage in
 the financial market of securities.
The modeled  uncertainty  of the financial market forces us to consider a weaker notion of arbitrage. 
\newline
 Let an event be $\mathcal{P}$-quasi surely true if it holds $P$-almost surely for each $P\in\mathcal{P}$.
\begin{defi}
Let  $\mathcal{R}\subset \mathcal{P}$. We say there is an $\mathcal{R}$-arbitrage opportunity in $\mathcal{M}(1,S,\mathcal{A})$ if there
exist  an admissible pair
$\eta\in\mathcal{A}$  such that  $\eta_0 S_0\leq 0$,
\begin{align*}
\eta_T S_T&\geq 0 ~~\mathcal{R}-quasi \:surely,\quad \text{and}\\
P\left(\eta_T S_T> 0\right)&>0 ~~\text{for at least one
}P\in\mathcal{R}.
\end{align*}
\end{defi}
The choice of the definition is based on the following observation. 
This arbitrage  strategy is   riskless for each $P\in\mathcal{R}$ and if the prior $P$ constitutes the
market one would gain a profit from  with positive probability.  With this  in mind, our $\mathcal{P}$-arbitrage notion can be seen as a  weak arbitrage of second order.
\newline
We say that a claim $X^m\in L^2({P})$ is \textit{marketed} in $\mathcal{M}(1,S,\mathcal{A})$ at time zero under $P\in\mathcal{P}$ if there is  a $\eta\in\mathcal{A}$ such that
$X^m= \eta_T S_T$ ${P}$-almost surely. In this case we say $\eta$ hedges $X^m$ and lies in $M_P$.  $\eta_0 S_0=\pi_P(X^m)$ is the price of $X^m$ in $\mathcal{M}(1,S,\mathcal{A})$ under $P\in\mathcal{P}$.
\newline
With Example 3 in mind, fix the   marketed spaces $M_P\subset L^2({P})$, $P\in\cal{P}$. The price of a marketed claim under the prior $P$ should to be well defined. Let $\eta,\eta'\in \mathcal{A}(P)$ generating the same claim $X^m\in M_P$, i.e. $\eta_T S_T={\eta'}_T S_T$ $P$-a.s., where $\mathcal{A}(P)$ refers to  self-financing portfolios under $P$. We have $\eta_0 S_0={\eta'}_0 S_0=\pi_P(X^m)$ under no $P$-arbitrage. Note, that this may not be true under no  $\hat P$-arbitrage, with $P\neq \hat P\in\mathcal{P}$. 
 This is related to the law of one price under a fixed prior. Now, similarly to the single prior case, we define viability in a financial market.
 We say that a financial market $\mathcal{M}(1,S,\mathcal{A})$ is \textit{viable} if it is $\Gamma(\mathcal{P})$-arbitrage free and the associated price system $(\{M_P,\pi_P\}_{P\in \mathcal{P}},\Gamma)$ is scenario-based viable.

\subsection{Extensions of price systems and EsMM sets}
In Section 2 we introduced the price space of sublinear price functionals generated by a set of linear
 $c_{2,\mathcal{P}}$-continuous functionals. The extension of the price functional is strongly related to the involved linear functionals
which constitutes the price systems locally. In this fashion, we introduce a modified notion of fair pricing. In essence, we associate  a risk neutral prior to each local and linear extension of a price system. Here, the term  local refers to a fixed prior, therefore no uncertainty is present.
\newline
In our uncertainty model, the price of a claim  equals the (discounted) value under a specific sublinear expectation. 
 Exploration of available information, when multiple priors are present, changes the view of a rational expectation.
\newline
In economic terms, the notion of symmetric martingales eliminates ambiguity in the valuation. It seems appropriate to introduce a  rational pricing principle of sublinear expectations.\footnote{The mutually singular priors generate a different view for the pricing  of a contingent claim.} This motivates the following definition.
\begin{defi}
 A set of probability measures $\mathcal{Q}$ on $(\Omega,\mathcal{B}(\Omega))$ is called an \textnormal{equivalent symmetric martingale measure set} (EsMM set) if the following two conditions hold:
\begin{enumerate}
\item For every $Q\in\mathcal{Q}$ there is a $P\in k(\mathcal{P})$ such that $P$ and $Q$ are equivalent to each other, such that $\frac{d P}{d Q}\in L^{2}({P})$.
\item The risky asset $S$ is a symmetric $\mathcal{E}^{\mathcal{Q}}$-martingale, where $\mathcal{E}^{\mathcal{Q}}$ is a sublinear expectation given by the supremum of expectations over $\mathcal{Q}$.
\end{enumerate}
\end{defi}
The first condition formulates a direct relation between an elements $Q$ in the  EsMM set $\mathcal{Q  }$ and the primitive priors $P\in\mathcal{P}$. The square integrability is a technical condition that guarantees the association to the
equilibrium theory of Section 2. The second  is the translated martingale condition.\footnote{It seems possible to proof that if the price process is not a symmetric martingale but a martingale  then arbitrage is possible. However, such considerations lies not in scope of this thesis.}
The rational expectation hypothesis and the idea of a fair gamble should 
establish maximal neutrality.  Under the new sublinear expectation the asset price and hence the portfolio process are symmetric martingales. This implies, as discussed in the introduction, that the value of the claim does not depend on the prior, i.e. the valuation is mean unambiguous.
\newline
The case of only one prior  is related to the well-known risk-neutral evaluation principle. 
Here, this principle needs a new requirement due to the more complex uncertainty model. In this sense the symmetry condition is responsible 
for the  \textit{uncertainty neutrality}.
\begin{rem}
 Note that in the case of a single prior framework, i.e. $\mathcal{P}=\{ P \}$, the notion of EsMM  sets  is reduced to accommodate equivalent martingale measures. In this regard we can think of  canonical generalization.
\newline
On the other hand, classical equivalent martingale measures (EMM) and a linear price theory are still present. Every single valued EsMM set $\mathcal{Q}=\{ Z_P \cdot P\}$ can be seen as an EMM under $P\in\mathcal{P}$. Here, the consolidation is given by $\Gamma=\delta_P$ and we have $\Gamma(\mathcal{P})=\{P\}$. 
\end{rem}
The following result motivates the discussion involving maximal risk neutrality and symmetry condition. The one to one mapping of Theorem 2
and hence the choice of the price space are appropriate. In this spirit we show that  $\mathcal{R}$-arbitrage in $\mathcal{A}$ with $\Gamma (\mathcal{P})=\mathcal{R}$  is inconsistent with an economic equilibrium for agents in $\mathbb{A}(\mathcal{P})$. We fix an associated price system  using procedure described at the end of Subsection 3.1.
\begin{thm}
Suppose the financial market model $\mathcal{M}(1,S,\mathcal{A})$ does not allow
any $\mathcal{P}$-arbitrage opportunity.
\newline
Then there is a bijection between EsMM-sets and sublinear   $\Psi:L^2(\mathcal{P})\rightarrow \mathbb{R}$ in $\mathfrak{P}(\mathcal{P})$ such that
$\Psi_{\restriction \mathbb{M}(\Gamma)}=\psi$ and $\Pi_{P\restriction M_P}=\pi_P$, $P\in\Gamma(\mathcal{P})$. 
The relationship is given by 
\begin{eqnarray*}
 \Psi(X)= \sup_{Q\in \mathcal{R}^*}E^{P}[X]=\mathcal{E}^{\mathcal{R}^*}( X),
\end{eqnarray*}
where $\mathcal{R}\subset k(\mathcal{P})$ and $\mathcal{R}^*=\{Z_P \cdot P: P\in \mathcal{R}, Z_P P=\Pi_P\}$  is an EsMM-set. 
\end{thm}
 Let  $\mathcal{R}\subset \mathcal{P}$ and $\mathfrak{M}(\mathcal{R})$ be the set of all  EsMM-sets $\mathcal{Q}$ such that the related consolidation $\Gamma$ satisfies $\Gamma(\mathcal{P})=\mathcal{R}$ . Theorem 2 can be seen as the  formulation of a one-to-one mapping between  
\begin{eqnarray*}
\mathfrak{P}(\mathcal{P}) \: \textnormal{ and  }\:\bigcup_{\mathcal{R}\subset k(\mathcal{P})}\mathfrak{M}(\mathcal{R}).
\end{eqnarray*}
 There is a hierarchy of sublinear expectation martingales, related to the chosen consolidation operator $\Gamma$ and the EsMM-set. We illustrate the relationship between $\Gamma$ and an EsMM-set in the following example.
\begin{ex}
We illustrate the relationship between  EsMM-sets and the consolidation operation $\Gamma$ when a price system is given.
For the sake of simplicity, let us assume that  $\{P_1,P_2,P_3,P_4\}=\mathcal{P}$. Starting with the sublinear price system, we have three price functionals $\pi_1,\pi_2,\pi_3$ and the consolidation operator $\Gamma$. Let us assume that $\Gamma=(+, \wedge)$ and $\lambda\in(0,1)$. This gives us
$\lambda\pi_1+ (1-\lambda)\pi_2=\pi^\lambda$ and  $\Gamma(\pi_1,\pi_2,\pi_3)=\pi^\lambda\wedge \pi_3$. The resulting  EsMM-set is given by $\mathcal{R}^*=\{Z^\lambda\cdot P^\lambda,Z_3P_3\}\in \mathfrak{M}( \mathcal{P}\setminus \{P_4 \})$, where  $P^\lambda=\lambda P_1+(1-\lambda)P_2$ and $Z^\lambda=\lambda Z_1+(1-\lambda)Z_2$.
\end{ex}
We close this consideration with some results analogous to those of the single prior setting where we combine   Theorem 2 and Theorem 1.  
\begin{cor}  Let $\mathcal{R}\subset \mathcal{P}$, such that $\mathcal{R}=\Gamma (\mathcal{P})$.
\begin{enumerate}
\item Scenario-based viability of $\mathcal{M}(1,S,\mathcal{A})$ is equivalent to the existence of an EsMM-set.
%\item We have $X\in \mathbb{M}(\mathcal{P})$ if and only if $\mathcal{E}^{\mathcal{Q}}(X)$ is a constant for every $\mathcal{Q}\in\mathfrak{M}(\mathcal{R})$.
\item Market completeness, i.e $M_P=L^2({P}) $ for each $P\in \mathcal{R}$, is equivalent to the existence of exactly one EsMM-set in $\mathfrak{M}(\mathcal{R})$.
\item If $\mathfrak{M}(\mathcal{R})$ is nonempty, then there is exists no $\mathcal{R}$-arbitrage.
\item If there is a strategy $\eta\in \mathcal{A}$ with $\eta_0 S_0$, $\eta_T S_T \geq 0$ $\mathcal{R}$-q.s. and $\mathcal{E}^\mathcal{R}(\eta_T S_T)>0$ then there is an $\mathcal{R}$-arbitrage opportunity. 
\end{enumerate}
\end{cor}
The result does not depend on the preference of the agent. The expected return under the sublinear expectation $\mathcal{E}^\mathcal{Q}$ equals the riskless asset. Hence, the value of a claim can be considered as the future value in the uncertainty-free world.\footnote{However, the sublinear expectation depends on $\Gamma$. }

\subsection{A special case: $G$-expectation}
Now, we select a stronger calculus to model the asset prices as a stochastic
 differential equation driven by a $G$-Brownian motion.\footnote{An illustration of the concept in a discrete time framework is achievable, via an application of the results in \cite{cnp11}.}
In this situation the volatility of the process concentrates the uncertainty in terms of the derivative of 
the quadratic variation. The quadratic variation of a $G$-Brownian motion creates uncertain volatility.
\newline
Again, we review the related result of the single prior framework.
\\[1em]
\textbf{Background:  It\^o processes in the single prior framework}
\newline
Now, we specify the asset price in terms of an It\^o process 
\begin{align*}
dS_t=\mu_t S_t dt+\sigma_t S_t dB_t,\quad S_0=1,
\end{align*}
  driven by a Brownian motion $B=\{B_t\}_{t\in[0,T]}$ on the given filtered probability space, $\mu,\sigma$ are processes such that  $S$ is a well defined processes in $\mathbb{R}_+$. The filtration  is generated by $B$. The interest rate is $r=0$.   Let $E^\theta$ be the exponential martingale, given by $d E^\theta_t= E^\theta_t \theta_t dB_t$, $E^\theta_0=1$,  with a Novikov consistent kernel $\theta$  we can apply Girsanov theorem. The following result  is from \cite{hakr79}:
\begin{quote}
The set of equivalent martingale measures is not empty if and only if 
$\rho=E^\theta_T\in L^2(P)$, $\theta\in L^2(P\otimes dt)$ and $S^*=\int \sigma dB$ is a $P$-martingale.
\end{quote}
$\rho$ can be interpreted as a state price density. The associated market price of risk $\theta_t=\frac{\mu_t-r}{\sigma_t}$ is the  Girsanov or pricing kernel of the state price density.

\subsubsection{Security prices as $G$-It\^o processes and sublinear valuation}
Our sublinear expectation is given by the  $G$-expectation $E_G:L^2(\mathcal{P})\rightarrow\mathbb{R}$.\footnote{It is shown in Theorem 52 of \cite{dhp11}, that this sublinear expectation can be represented by a weakly compact set, when the domain is in $L^1_G(\Omega)$.}
The construction of $ E_G$ on $L^2(\mathcal{P})$ can be achieved when the sublinear expectation space $(C([0,T];\mathbb{R}), \mathcal{C}_b(C([0,T];\mathbb{R})),E_G)$ as given, (see Appendix B.3 and references therein for more precise treatment).
\newline
 The   Girsanov theorem  is precisely what is needed
to verify  the  symmetric $G$-martingales property of the price processes $S$ under some sublinear expectation given by an EsMM-set. 
\newline
This uncertainty model enables us to apply the necessary stochastic calculus. As such, we model the financial market in the
 $G$-expectation setting, introduced in \cite{pe07;pe10}. Central results, such as a martingale representation,
a Girsanov type result, and a well behaved  underlying topology are desired for the foundational grounding of asset 
pricing. 
\newline
We select the  next rational base, namely an interval $[\sigma_1,\sigma_2]\subset \mathbb{R}_{++}$, instead of a  constant
 volatility $\sigma$, (see Example 4). The bounds of the interval is a model improving substitution with respect to
 constant $\sigma\in \mathbb{R}_{++}$.
We introduce  an asset price process driven by a $G$-Brownian motion $\{B^G_t\}_{t\in[0,T]}$. In  Appendix B.3 we present 
a small primer of the applied results. 
\newline
The asset price is driven by the following $G$-stochastic differential equation
 \begin{eqnarray*}
d S_t = \mu(t,S_t) d\langle B^G \rangle_t + V(t,S_t) d B^G_t,\quad t\in [0,T],\quad S_0=1.
\end{eqnarray*}
Let $\mu:[0,T]\times \Omega \times \mathbb{R}\rightarrow \mathbb{R}$ and $V:[0,T]\times \Omega \times \mathbb{R}\rightarrow \mathbb{R}_+$ be processes such that a unique solution exists.\footnote{We refer to Chapter 5 in  \cite{pe10} for existence results of G-SDE's.} Moreover, let $V(\cdot,x)$ be a strictly positive process for each $x\in\mathbb{R}_+$.
The riskless asset has interest rate zero.
\newline
The second condition of Definition 4 in Subsection 3.2 highlights how a  Girsanov transformation should  relate to a symmetric $G$-martingale and thus guarantee the
non emptiness of the concept.
For this purpose we define the related sublinear expectation generated by an EsMM-set, 
$\mathcal{Q}=\{ZP:P\in\mathcal{P}\}$ and $X\in L^2(\mathcal{P})$:
\begin{eqnarray*}
\sup_{Q \in \mathcal{Q}} E^{Q}[X ]= \mathcal{E}^ \mathcal{Q} (X )=E_G(ZX)
\end{eqnarray*}
Theorem 3 below justifies the choice of this shifted sublinear expectation when the asset price is restrained to  a symmetric 
martingale for an uncertainty-neutral  expectation.
\newline
 Let us consider an exponential martingale  $\mathtt{E}$ under the $G$-expectation,
 with a pricing kernel $\theta\in M^2_G(0,T)$, defined in Appendix B.3:
\begin{eqnarray*}
d\mathtt{E}^{\theta}_t=\mathtt{E}^{\theta}_t\theta(t,S_t)  d  B_t^G,\:\:\mathtt{E}^{\theta}_0=1
\end{eqnarray*}
By application of the results in Appendix B.3, we can write $ \mathtt{E}^{\theta}$ in explicit form 
\begin{eqnarray*}
 \mathtt{E}^{\theta}_t=\exp\bigg(-\frac{1}{2}\int_0^t \theta(s,S_s)^2 d\langle B^G \rangle_s-\int_0^t \theta(s,S_s) d B^G_s\bigg),\quad t\in[0,T].
\end{eqnarray*}
Let the pricing kernel  solve $ V(t,S_t)\theta(t,S_t)=\mu(t,S_t)$ for every $t\in[0,T]$ $\mathcal{P}$-quasi surely.
Before we formulate the last result we define
\begin{eqnarray*}
S_t^*=S^*_0+\int_0^t V(s, S^*_s) dB_s^G,\quad  t\in[0,T]
\end{eqnarray*}
such that a unique solution on $(\Omega,\mathcal{H},E_G)$ exists, see \cite{pe10}.
\begin{thm}
The  set of $\mathcal{M}(\mathcal{P})$  of EsMM-sets is not only the empty set if and only if 
 $S^*$ is an $E_G$-martingale and
\begin{align*}
 E_G \bigg[\exp \bigg(\delta\cdot \int_0^T \theta^2_s d\langle B^G\rangle_s \bigg)\ \bigg]< \infty,
\end{align*} 
for some $\delta>\frac{1}{2}$.
\end{thm}
With Theorem 2 in mind we can include scenario based viability. Let $X\in L^2(\mathcal{P})$ be a contingent claim, such that it is  priced by $\mathcal{P}$-arbitrage with value  $\Psi(X)=E_G(\mathtt{E}^{\theta}_T X)$, whenever $\Gamma$ consists only of a consolidation via the  maximum operation. 
\begin{rem}
The more precise calculus of the $G$-expectation is based on a description of nonlinear partial differential equations. This allows us to  create a uniform state price density process in terms of an exponential martingale, based on a $G$-martingale representation theorem, (see Appendix B 3).
\newline
With this in mind,  a more elaborated notion of EsMM-sets can be formulated  by requiring that the densities $Z_P$, $P\in\mathcal{P}$ creates a uniform process as a symmetric martingale under sublinear expectation  $\mathcal{P}$.
\end{rem}
Extensions to continuous trading strategies seem straight forward. Nevertheless, an admissibility condition should be requested, in order to exclude doubling strategies. Considering markets with more than one uncertain security requires a multidimensional Girsanov theorem.\footnote{See \cite{os11}.}
\newline
Let us close this subsection with an example on  the connection between  superreplication of claims and EsMM-sets.
\begin{ex}
 Under one prior $P$, \cite{del92} obtained the superreplication price in terms of martingale measures in $\mathfrak{M}(\{P\})$:
\begin{eqnarray*}
\Lambda (X,P)& =&\inf \{y\geq 0|\exists~\theta\in\mathcal{A}:y+\theta_T S_T\geq X\:P-a.s.\}\\
&=& \sup_{Q\in  \mathfrak{M}(\{P\})} E^Q[X]
\end{eqnarray*}
When the uncertainty is given by a set of mutually singular priors,  a super-replication price can be derived, see  \cite{dema06}.
\begin{align*}
\Lambda (X,\mathcal{P}) =\inf \{y\geq 0|\exists~\theta\in\mathcal{A}(\mathcal{P}):~y+\theta_T S_T\geq X~~\mathcal{P}-q.s.\}
\end{align*}
 in terms of an \textit{unknown} set of martingale laws $ \mathcal{M}$ such that
\begin{align*}
\Lambda (X,\mathcal{P}) =\sup_{Q\in \mathcal{M}} E^Q[X].
\end{align*} 
In turns out that in the $G$-framework with simple trading strategies  this  set is an EsMM-set.When applying our theory to this problem, we get
\begin{align*}
\Lambda (X,\mathcal{P}) =\sup_{P\in \mathcal{P}} \sup_{Q\in\mathfrak{M}(\{P\})} E^Q[X]= E_G[\mathtt{E}_T X],
\end{align*} 
uppn applying  Theorem 3 and Theorem 3.6 of \cite{vor10}
and is  associated to the maximal EsMM-set in $\mathfrak{M}(\mathcal{P})$. However, an easy consequence is 
that every EsMM-set delivers a price below the superhedging price.
\end{ex}

\section{Discussion and Conclusion}
We present a  framework  and a theory of   derivative security pricing  where the  uncertainty model is given by a set of singular probability measures which incorporate volatility uncertainty. The notion of equivalent martingale measures changes, and the related linear expectation principle becomes a sublinear theory of valuation. The associated arbitrage principle should consider all remaining uncertainty in the consolidation.
\newline
The results of this paper may serve as a starting point to obtain a fundamental theorem of asset pricing (FTAP) under mutually singular uncertainty. In  \cite{desc94} and \cite{desc98},  this is achieved for the single prior uncertainty model in great generality. The notion of arbitrage is in principle a  separation property of convex sets in a topological space. In this regard, the choice of the underlying topological structure is essential for observing a FTAP. For instance, \cite{lesc95}, establish a FTAP with an \textit{approximate arbitrage} based on a different notion of convergence. 
\newline
As mentioned in the introduction, \cite{jou95} considered security markets with bid-ask spreads and introduced a modified notion of equivalent martingale measures. In this context a FTAP under transaction cost was proved in discrete time by \cite{sch04}, and  in continuous time by \cite{grs10}.
\newline
 In our setting, two aspects must be kept in mind for deriving a FTAP with mutually singular uncertainty.
\newline
Firstly, the spaces of claims and portfolio processes are based on a capacity norm, and thus forces one to argue for  the quasi sure analysis, a fact  implied in our definition of arbitrage, (see Definition 3). A corresponding notion of free lunch with vanishing \textit{uncertainty} will have to  incorporate  this more sensitive notion of random variables.
\newline
Secondly, the sublinear structure of the  price system allows for a nonlinear separation of convex sets. With one prior, the equivalent martingale measure separates achievable claims with arbitrage strategies. In our small meshed structure of random variables this separation is guided by the consolidation operator $\Gamma$.
\newline
Our preference-free pricing principle gives us a valuation via expected payoffs of different adjusted priors. In comparison to the  preference and  distribution free results in a perfectly competitive market, see \cite{ro76}, the implicit assumption is the common knowledge of uncertainty, described by a single probability measure. The  uncertainty preface dramatically  dictates the consequences for  pricing without a utility gradient approach of consumption-based pricing.
\newline
The valuation of claims, determined by $\mathcal{P}$-arbitrage, contains a new object $\Gamma$, which  may inspire skepticism. However, note that the consolidation  operator $\Gamma$  should  be seen as a too to regulate  financial markets. The valuation of claims in the balance sheet of a bank should depend on  $\Gamma$.  For instance, this may affect  fluctuations of opinion in the market as a consequence of uncertainty.
 In Remark 1 of Section 2 we describe how a good  consolidation may be found via consideration of mechanism design. Such considerations may provide  a base for the choice of the  valuation principle  under multiple priors.\footnote{A starting point could be \cite{lrs09}.} As a first heuristic, it is possible that utilitarian (convex combination) and Rawlsian  (supremum operation)  welfare functions may constitute a principle of fair pricing.
\newline
Before we close this paper with a discussion on asset pricing under uncertainty and an alternative interpretation of sublinear pricing, we state a technical comment.  The suborder dual $\mathfrak{P}(\mathcal{P})$ of Subsection 2.3 can be  elaborated using results on  embedding duals in the sub order dual, and could be useful for answer continuity questions about the Riesz-Kantorovich functional.
\\[1em]
\textbf{Preferences and Asset pricing}
\newline
The uncertainty  model in our paper is closely related to  \cite{epwa94} and \cite{epji11} as they consider equilibria with linear prices in their economy. This leads to an indeterminacy in terms of a continuum of linear equilibrium price systems. The relationship between uncertainty and indeterminacy is caused  by the constraint to pick one \textit{effective} prior. The Lucas critique\footnote{See Section 3.2 in \cite{epsc10}.} applies insofar as it describes the unsuitable usage of a pessimistic investor to fix an effective prior in reduced form. Our approach takes a \textit{preference free} view. We  value contingent claims in terms of  mean unambiguous asset price processes. In other words,   the priors of the uncertainty neutral  model yield expectations of the security price  that do \textit{not} depend on the chosen "risk neutral" prior.  Nevertheless, the idea of a risk neutral valuation principle is not appropriate, as  different mutually singular priors delivers different expectations, that cannot be related via a density.
\newline
 From this point of view, we disarrange the indeterminacy  of sublinear prices, and allow for the appearance of a planner to configure the sublinearity. In this sense, the regulator as a policy maker is now able to confront unmeasurable sudden fluctuations in the volatility. A single prior, as a part of the equilibrium output, can create an invisible threat of convention, which may be used to create the illusion of security when faced with an uncertain future. 
 In a model with mutually singular priors, the focus on a single prior creates a hazard. Events with a positive probability under an ignored prior may be a null set under an effective prior in a consumption-based view.
\\[1em]
\textbf{Sublinear prices and regulation via consolidation}
\newline
In this context, sublinearity is associated with the principle of
diversification. In these terms, equilibrium  with  a sublinear price system covers the concept of Walrasian prices which decentralize with the coincidental awareness of different scenarios. 
 A priori, the instructed Walrasian auctioneer has no knowledge of which prior $P$ in $\mathcal{P}$ occurs. The auctioneer assigns to each prior $P\in \mathcal{P}$ a locally  linear price $\pi_P$. 
 The degree of discrimination is related to the intensity of nonlinearity.
Note that this is a normative category and opens the door to   the economic basis of regulation.
Each prior is  a probabilistic scenario.The auctioneer \textit{consolidates} the price for each possible scenario into one certain valuation. This is also true for an  agent in the model, hence the auctioneer should be able to \textit{discriminate under-diversification} in terms of ignorance of priors in this uncertainty model. Further, a von Neumann-Morgenstern utility assumption result in an overconfidence of certainty in the associated agent.
\newline
Since we want to generalize fundamental theorems of asset pricing, we are concerned with the relationship between 
 equivalent martingale measures, viable  price systems, and  arbitrage. In this setting  these concepts must be recast in terms of the multiple prior uncertainty of the model. In contrast, with one prior an equivalent martingale measure is associated with  a linear price system. The underlying neoclassical  equilibrium concept is a fully positive theory.  In the multiple prior setting such
a price extension can be regarded as a diversification-neutral valuation principle. Here, diversification  refers to a  given  set of priors $\mathcal{P}$. Should  the unlucky situation arise that  an unconsidered prior
governs the market,   it is the task of the regulator to robustify these option via an appropriate price system.
For instance, uniting two valuations of contingent claims  cannot be  worse than adding the
two uncertain outcomes separately. This is the diversification principle under $\mathcal{P}$.

Recalling the quotation of  \cite{aty00}  in the introduction, the degree of  sublinearity in our  approximation is regulated by the type of consolidation of scenario-dependent linear price systems. These price systems act locally on each scenario $P\in\mathcal{P}$ in a linear fashion.

\begin{appendix}

\appendix

\section{Appendix: Details and Proofs}

\subsection{ Section 2}

\subsubsection{Details for Section 2}
Let $L^2(\mathcal{P})=\mathcal{L}^2(\mathcal{P}) /\mathcal{N}$ be the quotient of $\mathcal{L}^2(\mathcal{P})$ by the $c_{2,\mathcal{P}}$ null elements. Such null elements are characterized by random variables which are  $\mathcal{P}$-polar.  $\mathcal{P}$-polar sets  evaluated under every prior are zero or one. But the value may differ between different priors. A property holds \textit{quasi-surely} (q.s.) if it holds outside a polar set. 
Furthermore, the space $L^2(\Omega)$ is characterized by
\begin{eqnarray*}
L^p(\mathcal{P})=\{X\in
L^0(\Omega):X\text{ has a q.c. version, }
\lim_{n\rightarrow\infty}\mathcal{E}^\mathcal{P}(\vert X\vert)^21_{\{\vert X\vert >n\}}]=0\},
\end{eqnarray*}
where $L^0(\Omega)$ denotes the space of all measurable
real-valued functions on $\Omega$. A mapping $X:\Omega\rightarrow\mathbb{R}$ is said to be
quasi-continuous  if $\forall \epsilon
> 0$  there exists an open set $O$ with $\sup_{P\in\mathcal{P}}P(O) < \epsilon$ such
that $X|_{O^c}$ is continuous. r We say that $X:\Omega
\rightarrow \mathbb{R}$ has a \textit{quasi-continuous version} (q.c.) if
there exists a quasi--continuous function
$Y:\Omega\rightarrow\mathbb{R}$ with $X = Y$ q.s.
 The mathematical framework provided enables the analysis of
stochastic processes for several mutually singular probability
measures simultaneously. All equations are  understood in the sense of quasi-sure.
This means that a property holds almost-surely for all
 scenarios $P\in\mathcal{P}$.
\newline
When recast the order relation taken from \cite{bnk10} we have: $X\geq 0$ if and only if
\begin{eqnarray*} 
 &&\textnormal{there is a sequence } \{X_n\}_{n\in\mathbb{N}}\subset \mathcal{C}_b(\Omega), X_n\geq 0 \textnormal{ such that }\\
&& \forall Y\in L^2(\mathcal{P})\textnormal{ of class } X \textnormal { we have } \lim_{n\rightarrow \infty} c_{2,\mathcal{P}}(Y-X_n)=0.
\end{eqnarray*}
Since, for all $X,Y\in L^2(\mathcal{P})$ with $\vert X\vert\leq \vert Y\vert$ 	imply $c_{2,\mathcal{P}}(X)\leq c_{2,\mathcal{P}}(Y)$,
 we have that $L^2(\mathcal{P})$ is a Banach lattice.\footnote{This is of interest for existence result of general equilibria.}
\newline
Following we discuss the different operations for consolidation. Let $\Pi_P=ZP \in L^2(\mathcal{P})^*$, with $P\in\mathcal{P}$.
\newline  Let $ \mu$ be a measure on the Borel measurable space $(\mathcal{P},\mathcal{B}(\mathcal{P}))$    with $\mu(\mathcal{P})=1 $ and full support on $\mathcal{P}$. 
 In this context we can consider the  additive case in $\mathfrak{P}(\mathcal{P})$, where a new prior is generated:\footnote{The related operation of convex functionals would corresponds to the convolution operation. Since we have no assumption on the convexity of $\mathcal{P}$, the prior $P_{\mu}$ may only lie in the convex hull of  $\mathcal{P}$.}
\begin{eqnarray*}
 \Gamma_\mu :\prod_{P\in \mathcal{P}}L^2(\mathcal{P})^*\rightarrow \mathfrak{P}(\mathcal{P}),\quad \Gamma_\mu(\{\Pi_P\}_{P\in\cal P})=\int_{\mathcal{P}}E^P[Z \cdot] d\mu(P)=E^{P_{\mu}}[ Z \cdot]
\end{eqnarray*}
We can consider the Dirac measure as an example.  The related consideration of only one special prior in $\mathcal{P}$ is in essence the uncertainty model in \cite{hakr79}. The  operation in question  is given by  $ (\Pi_P)_{P\in\cal P}\mapsto E^{P}[Z\cdot]$.\footnote{A different point of view is that this case can be seen as a special case of \cite{rie11}.}
\newline
The second operation in $\mathfrak{P}(\mathcal{P}) $ is a point-wise maximum:
\begin{eqnarray*}
 \Gamma_{\sup} :\prod_{P\in \mathcal{P}} L^2(\mathcal{P})^* \rightarrow \mathfrak{P}(\mathcal{P}),\quad  \Gamma_{\sup} (\{\Pi_P\}_{P\in\mathcal{ P}}) =\sup_{P\in\mathcal{P}} E^{P}[Z\cdot]=\mathcal{E}^\mathcal{P}( Z\cdot).
\end{eqnarray*}
 This is an extreme form of consolidation and can be considered as the highest awareness of all priors.
Note that combinations between the maximum and addition operation are possible as indicated in Example 2 and Proposition 1.

\subsubsection{Proofs of Section 2} 
\begin{ppo}
 The 5th claim follows from Theorem 1 in \cite{bifi10}, whereas the other claims follow directly from the construction of the functionals in $\mathfrak{P}(\mathcal{P})$.
.  $\hfill\blacksquare$
\end{ppo}
For the proof of Theorem 1, we define the shifted preference  relationship $\succsim^0$ such that every feasible net trade is worse off than $(0,0)\in B(0,0,\psi,\mathbb{M}(\Gamma)) $.
Obviously, an agent given by $\succsim^0$ does not trade. Hence, an initial endowment constitutes a no trade equilibrium. 
\begin{tpo}
 Let the price system $(\{M_P,\pi_P\}_{P\in \mathcal{P}},\Gamma)$ be given and we have a $\Psi\in \mathfrak{P}(\mathcal{P})$ on $L^2(\mathcal{P})$ such that $\Psi_{\restriction\mathbb{M}(\Gamma)}=\psi$. The relation on $\mathbb{R}\times L^2(\mathcal{P})$, given by
\begin{eqnarray*}
 (x,X) \succcurlyeq (x', X')\quad if \:\: x+-\Psi(-X)\geq x'+-\Psi(-X'),
\end{eqnarray*}
is an element  of $\mathbb{A}(\mathcal{P})$. This can be justified by the $c_{2,\mathcal{P}}$-continuity and concavity of $-\Psi(-\cdot)$.
 The bundle $(\hat x,\hat X)=(0,0)$ satisfies 
the viability condition of Definition 2, hence $(\{M_P,\pi_P\}_{P\in \mathcal{P}},\Gamma)$ is scenario-based viable.
\newline
In the other direction, let  $\psi:\mathbb{M}(\Gamma)\rightarrow \mathbb{R}$ be a price system, $\succcurlyeq\in\mathbb{A}(\mathcal{P})$, 
 and  $(\hat x,\hat X)$ satisfy the viability condition. We may assume
$(\hat x,\hat X)=(0,0)$,  since  it  is only a geometric deferment. Consider the following sets
\begin{eqnarray*}
\succ^0&=&\{(x,X) \in\mathbb{R}\times L^2(\mathcal{P}): (x,X)\succ (0,0)  \},\\\\
B(0,0,\psi,\mathbb{M}(\Gamma))&=&\{(x,X^m)\in\mathbb{R}\times \mathbb{M}(\Gamma) : x +\psi(X^m)\leq 0  \}.
\end{eqnarray*}
We have that $B(0,0,\psi,\mathbb{M}(\Gamma))$ and  $\succ^0$ are convex sets. By the $c_{2,\mathcal{P}}$-upper semi continuity of $\succsim$ ,  $\succ^0$ 
is  $c_{2,\mathcal{P}}$-open. We apply  Lemma 1, stated below in terms of a nonlinear
 separation theorem. A non zero $c_{2,\mathcal{P}}$-continuous sublinear  functional 
on $\mathbb{R}\times L^2(\mathcal{P})$ with $\phi(x,X)\geq 0$ for all $(x,X)\in \succ^0$ and
 $\phi(x,X)\leq 0$ for all $(x,X)\in  B(0,0,\psi,\mathbb{M}(\Gamma))$ are constructed.	
\newline
  There is a $(y,Y)$ with $\phi(y,Y)<0$, since $\phi$ is non trivial.  Strict monotonicity implies 
$(1,0)\succ (0,0)$. The continuity  $\succcurlyeq$ gives us $(1+\epsilon y,\epsilon Y)\succ(0,0)$, 
for some $\epsilon>0$, hence 
\begin{eqnarray*}
 -\phi(1+\epsilon x',\epsilon X')&=& -\phi(1,0)+\epsilon \phi(y,Y)\leq 0  \\
and\:\:\phi(1,0)&\geq &-\epsilon \phi(y,Y) >0
\end{eqnarray*}
We have $\phi(1,0)>0$ and after a renormalization let $\phi(1,0)=1$. Moreover write $\phi(x,X)=x+\Psi(X)>0$, where $\Psi:L^2(\mathcal{P})\rightarrow \mathbb{R}$ is a functional in $\mathfrak{P}(\mathcal{P})$.
\newline
Strict positivity of $\Psi$ follows from $(0,x)\succ(0,0)$, hence $(-\epsilon,x)\succ(0,0)$, and therefor
$\Psi(x)-\epsilon\geq 0$.
\newline
Let $X^m\in\mathbb{M}(\Gamma)$, since $(-\psi(X^m),X^m),(\psi(X^m),-X^m)\in B(0,0,\psi,\mathbb{M}(\Gamma))$ we have $0=\phi(\psi(X^m),X^m)
=\psi(X^m)-\Psi(X^m)$ and $\Psi_{\restriction \mathbb{M}(\Gamma)}= \psi$ follows. $\hfill\blacksquare$
\end{tpo}
The following lemma is applied  to the proof of Theorem 1. Let $\mathfrak{P}_{ \restriction \mathbb{M}(\Gamma)}(\mathcal{P})$ be the space a of all functionals $\psi\in\mathfrak{P}(\mathcal{P})$ with domain $\mathbb{M}(\Gamma)$.
\begin{lem}
Let $\psi\in\mathfrak{P}_{\restriction\mathbb{M}(\Gamma)}(\mathcal{P})$ then there is a $\Psi\in\mathfrak{P}(\mathcal{P})$ with
$\Psi_{\restriction \mathbb{M}(\Gamma)}= \psi$.
\end{lem}
Note, that this is a  Hahn Banach type result for functionals in $\mathfrak{P}(\mathcal{P})$. 
We illustrate this in  the following diagram: 
\\[1em]
\begin{xy}

 \qquad  \qquad \xymatrix{
      \{\pi_P:M_P\rightarrow \mathbb{R} \}_{P\in \mathcal{P}} \ar[r]^{\Gamma} \ar[d]_{Hahn\: Banach}    &   \psi:\mathbb{M}(\Gamma)\rightarrow \mathbb{R} \ar[d]  \\
\{\Pi_P:L^2(\mathcal{P})\rightarrow \mathbb{R} \}_{P\in \mathcal{P}}   \ar[r]_{\quad\Gamma}             &   \Psi  :L^2(\mathcal{P})\rightarrow \mathbb{R} 
  }

\end{xy}
\begin{lpo} Fix  $ \psi:\mathbb{M}(\Gamma) \rightarrow\mathbb{R}$, given by $\Gamma(\{\pi_P\}_{P\in \mathcal{P}})$.
We apply Hahn Banach for each $P\in \mathcal{P}$ with respect to  $\pi_P : M_P \rightarrow\mathbb{R}$. We have a collection of 
$\Pi_P:L^2(\mathcal{P}) \rightarrow\mathbb{R}$ such that $\Pi_{\restriction M_P}=\pi_P$. Hence, $\Psi =\Gamma(\{\Pi_P\}_{P\in\mathcal{P}})$. 
By the definition of the price space, we have $\Psi\in\mathfrak{P}(\mathcal{P})$. $\hfill\blacksquare$
\end{lpo}

\subsection{Section 3}

\subsubsection{Details  of Section 3}
Next, we discuss the augmentation of our information structure. The unaugmented filtration  is given by $\mathbb{F}^o$.
As mentioned in Subsection 3.1, the set of priors must be stable under pasting, in order to apply the framework of \cite{nuso10}. For the sake of completeness we recall this notion.
\begin{defi}
The set of priors is \textnormal{stable under pasting} if for every $P\in\mathcal{P}$,      every $\mathbb{F}^o$-stopping time $\tau$, $B \in \mathcal{F}_\tau^o$ and $P_1,P_2\in \mathcal{P}(\mathcal{F}_{\tau}^o,P)$, the prior $P_\tau$ given by
\begin{align*}
P_\tau= E^P\big[P_1(A\vert \mathcal{F}_\tau^o ) 1_B + P_2 (A\vert \mathcal{F}_\tau^o)1_{B^c}\big], \quad A\in \mathcal{F}_\tau^o
\end{align*}
is a prior in $\mathcal{P}$.
\end{defi}
In the multiple prior setting, with a given reference measure this property is equivalent to the well known notion of \textit{time consistency}. However, this is not true if there is no dominant prior. Additionally, the set of priors  must be chosen maximally. This is a property which holds  for a \textit{fixed} set of random variables. For further consideration, we refer the reader to Section 3 in \cite{nuso10}.
\newline
The usual condition of a "rich" $\sigma$-algebra at time $0$ is widely used in mathematical finance. But the economic meaning is questionable. Our uncertainty model  of mutually singular priors  can be augmented, similarly to the classical case, using the right continuous filtration given by 
$\mathbb{F}^+=\{\mathcal{F}^+_t\}_{t\in[0,T]}$ where 
\begin{align*}
\mathcal{F}^+_t=\bigcap_{s>t}\mathcal{F}^o_t,\:\textnormal{for} \:\:t\in[0,T[
\end{align*}
The second step is to augment the minimal right continuous filtration $\mathbb{F}^+$ by  all polar sets of $(\mathcal{P},\mathcal{F}^o_T)$, i.e. $\mathcal{F}_t=\mathcal{F}_t^+\vee\mathcal{N}(\mathcal{P},\mathcal{F}^o_T)$. This augmentation is strictly smaller than the universal augmentation $\bigcap_{P\mathcal{P}} \overline{\mathbb{F}^o}^P$. This choice is economically reasonable as the initial $\sigma$-field contains not all 0-1  limit events. An agent considers this  exogenously
specified information structure. It describes what information the agent  \textit{can} know at each date. This is the analogue to a filtration in the single prior framework satisfying the usual conditions. 
\newline
 According to  Appendix B.2,   we have  a countable set  $\{P_n\}_{n\in\mathbb{N}} \subset \mathcal{P}$ 
such that for every  positive random variable $X$ in $L^2(\mathcal{P})$ we have
\begin{eqnarray}
 \mathcal{E}^{\mathcal{P}}(X)=\sup_{n\in \mathbb{N}} E^{P_n}[X].
\end{eqnarray}
We write $(P_n)\sim \mathcal{P}$ for  a  set, which allows such a countable reduction.\footnote{This reduction  is heavily related to the weak compactness of $\mathcal{P}$, see \cite{bnk10}. } Note that the related
Banach spaces are the same, see \cite{bnk10}.

The $\mathcal{P}$-arbitrage condition
 can be reformulated with a special prior $\tilde P$ in a  simpler form: $\eta_T S_T \geq 0  \:\:\tilde P-a.s$ and $\tilde P(\eta_T S_T> 0)>0 $. 
Without convexity of $\mathcal{P}$, $\hat P\in\mathcal{P}$ is not necessarily true.\footnote{Note that the arbitrage definition has only positive random variables under consideration. This allows us to consider an arbitrage controlling prior in the canonical class, see Appemdix B.2.}

\subsubsection{Proofs of Section 3}

For the proof of Theorem 2 we need a result from Appendix B.1.  We formulate a generalized Riesz representation Theorem: A linear functional  $\Pi$ on $L^2(\mathcal{P})$
is  $c_{2,\mathcal{P}}$-continuous  if and only if for every $X\in L^2(\mathcal{P})$ we have  $\Pi(X)=E^P[Z_P X]$ for some $P\in \mathcal{P} $ and $Z_P\in L^2(P)$.

\begin{tpo}
We fix an  EsMM-set $\mathcal{Q}$. The related consolidation $\Gamma$ gives us the set of relevant priors $\Gamma(\mathcal{P})\subset \mathcal{P}$. Let $Z_P= \frac{d P}{dQ}$, for each $Q\in \mathcal{Q} $ and the related $P\in \mathcal{P} $.
We have  $Z_P\in L^{2}({P})$ Let  a strictly positive $\Psi\in \mathfrak{P}(\mathcal{P})_{++}$ be given.
\newline 
 Take a marketed claim $X^m\in \mathbb{M}(\Gamma)$ and let $\eta\in \mathcal{A}$ be a self-financing trading strategy
that hedges $X^m$. This gives us the following equalities, since $\eta\in \mathcal{A}$, by the rule for conditional $\mathcal{E}$-expectation and since $S$ is a symmetric $\mathcal{E}^{\mathcal{Q}}$-martingale,  $0\leq t\leq u\leq T$, 
\begin{eqnarray*}
 \mathcal{E}^{A^*}_u(\eta_t S_t)=\eta_t^+ \mathcal{E}^{A^*}_u( S_t)+\eta_t^- \mathcal{E}^{A ^*}_u( -S_t)=\eta_t^+  S_u -\eta_t^- S_u=  \eta_u  S_u,
\end{eqnarray*}
where $\eta=\eta^+- \eta^-$ with $\eta^,- \eta^-\geq 0$ $\mathcal{P}$-quasi surely. Therefore we achieve
 \begin{eqnarray*}
 \Psi(X^m)= \mathcal{E}^{A*}_0(\eta_T S_T)=\eta_0 S_0=\psi(X^m).
 \end{eqnarray*}
For the other direction let $\Psi\in \mathfrak{P}(\mathcal{P})_{++}$ with $\Psi_{\restriction  \mathbb{M}(\Gamma)}=\psi$, related to a  set of linear functionals $\{\pi_P:M_P\rightarrow \mathbb{R}\}_{P\in\mathcal{P}}$ and $\{\Pi_P:L^2(\mathcal{P})\rightarrow \mathbb{R}\}_{P\in\mathcal{P}}$, such that $\Pi_{\restriction  M_P}=\pi_P$. Now, we define $\mathcal{Q}$ in terms of $\Gamma$.
\newline
We discuss the possible cases which can appear. For simplicity we assume $\mathcal{P}=\{P_1,P_2,P_3\}$
. Let $P^{k,j}=\frac{1}{2} P^k+ \frac{1}{2} P^j$ and $Z^{k,j}=\frac{1}{2} Z^k+ \frac{1}{2} Z^j$, recall that we can represent each functional $\Pi_P$ by $Z_P P$.
\begin{enumerate}
\item $\frac{1}{2}\Pi_1 +\frac{1}{2} (\Pi_2 \wedge \Pi_3)$ becomes $\{Z^{1,2} P^{1,2}, Z_3 P_{3}\}=\mathcal{Q}$
\item $(\frac{1}{2}\Pi_1 + \frac{1}{2}\Pi_2 )\wedge \Pi_3$ becomes  $  \{Z_1 P_1, Z^{2,3} P^{2,3}\}=\mathcal{Q}$
\end{enumerate}
Since $\mathcal{Q}=\{P Z_P: P\in \Gamma(\mathcal{P}),Z_P\in L^2(P)\}$, the first   condition of Definition 4 follows, note that the square integrability of each $Z_P$ follows from the $c_{2,\mathcal{P}}$-continuity of linear functionals which generate $\Psi$.
\newline 
We prove the symmetric martingale property of the asset price process. Let $B\in \mathcal{F}_t$,  $\eta\in\mathcal{A}$ be a self-financing trading strategy and
\begin{eqnarray*}
 \eta_{s}=\begin{cases}
  1  & s\in[t,u[\text{ and } \omega\in B\\
0 &\text{ else },
\end{cases}
\quad
 \eta^0_{s}=\begin{cases}
  S_t,  & s\in[t,u[\text{ and } \omega\in B\\
 S_u-S_t, &s\in[u,T[\text{ and } \omega\in B\\
0 &\text{ else}.
\end{cases}
\end{eqnarray*}
This strategy yields a portfolio value
\begin{eqnarray*}
 \eta_T S_T=(S_u-S_t)\cdot 1_B,
\end{eqnarray*}
the claim $\eta_T S_T$ is marketed at price zero. In terms of the modified sublinear expectation 
$ \{\mathcal{E}_t^{\mathcal{Q}}(\cdot)\}_{t\in[0,T]}$, we have with $t\leq u$
\begin{eqnarray*}
 \mathcal{E}_t^{\mathcal{Q}}((S_t-S_u)1_B)=0
\end{eqnarray*}
By Theorem 4.7 \cite{10}, it follows that $S_u= \mathcal{E}_t^{\mathcal{Q}}(S_u)$.\footnote{The result is proven for the $G$-framework. However the assertion is in our setting true as well, by an application of   Theorem 4.10 of \cite{nuso10} instead of  Theorem 4.1.42 of \cite{pe10}.} But this means that $\{S_t\}_{t\in[0,T]}$  is   $\mathcal{E}^{\mathcal{Q}}$-martingale. The same argumentation holds for $-S$, hence the asset price $S$ is  a symmetric  $\mathcal{E}^{\mathcal{Q}}$-martingale. $\hfill\blacksquare$
\end{tpo}

\begin{cpo}
\begin{enumerate}
\item Suppose there is a $\mathcal{Q}\in \mathcal{M}(\mathcal{P})$ and let $\eta\in \mathcal{A}$ such that $\eta_TS_T\geq 0$ and $ P(\eta_TS_T> 0)>0$ for some $P\in\mathcal{P}$. Since for all $Q\in\mathcal{Q}$ there is a $P\in k( \mathcal{P})$ such that $Q\sim P$, there is a $Q'\in \mathcal{Q}$ with $Q'(\eta_TS_T>0)>0$. Hence, $\mathcal{E}^\mathcal{Q}(\eta_T S_T)>0$ and by Theorem 2 we observe $\mathcal{E}^\mathcal{Q}(\eta_T S_T)=\eta_0 S_0 $. This implies that no $\mathcal{P}$-arbitrage exists.
%\item According to Corollary 4.11 in \cite{nuso10} the mean unambiguity  of  $X$ is equivalent to being that $X$ is replicable, i.e. 
%\begin{eqnarray*}
%X=\int_0^T H_s d B_s,\quad for every\:\: P\in\mathcal{R},
%\end{eqnarray*}
%where $\int_0^T  \vert H_s\vert^2 d\langle B \rangle_s<\infty$ $\mathcal{R}$-q.s.
 % Here, the stochastic integral is defined $P$-a.s. By the self financing %property we have $\eta_T S_T$ 
\item  In terms of Theorem 1, each $P\in \mathcal{R}$ admits exactly one extension. With Theorem 2 the result follows.
\item  By Theorem 2 this is equivalent to  the non emptiness of $\mathfrak{P}(\mathcal{P})$ . Fix a $\Psi\in \mathfrak{P}_{++} (\mathcal{P})$, with $\Gamma(\mathcal{P})=\mathcal{R}$ and a $\eta \in \mathcal{A}$ such that $\eta_0 S_0=0$ hence $\Psi(\eta_T S_T)=0$. The viability of $\Psi$ implies $\eta_T S_T=0$   $\mathcal{R}$-q.s. Hence, no $\mathcal{R}$-arbitrage exist.
\item This then follows by the same argument as in \cite{hapl81} (see the Lemma on p.228).$\hfill\blacksquare$
\end{enumerate}
\end{cpo}

For the  proof of Theorem 3, we apply results from  stochastic analysis in the $G$ framework. The results are collected in Appendix B.3.
\begin{tpo}
In accordance to Remark 3, let $\mathcal{Q}$ be an EsMM-set, given by  $\mathcal{Q}= \{\rho P: P\in\mathcal{P}\}$, where  the density $\rho$ with  $\rho\in L^2(\mathcal{P})$ and  ${E}_G[\rho]=-{E}_G[-\rho]$. Next  define  the stochastic process $(\rho_t)_{t\in[0,T]}$ by $\rho_t=E_G[\rho\vert \mathcal{F}_t]$
resulting in a symmetric $G$-martingale to which we apply the martingale representation theorem for $G$-expectation, stated in Appendix B.3.
\newline
Hence, there is a $\gamma\in M^2_G(0,T)$ such that  we can write
\begin{eqnarray*}
 \rho_t= 1+\int_0^t \gamma_s dB^G_s, \quad t\in[0,T],\quad  \mathcal{P}-q.s.
\end{eqnarray*}
By the G-It\^o formula, stated in the Appendix B.3, we have 
\begin{eqnarray*}
ln(\rho_t)= \int_0^t \phi_s dB^G_s+\frac{1}{2} \int_0^t \phi^2_s d\langle B^G\rangle_s,\quad  \mathcal{P}-q.s
\end{eqnarray*}
for every  $ t\in[0,T]$ in $L^2_G(\Omega_t)$ and hence
\begin{eqnarray*}
 \rho=\mathtt{E}^{\phi}_T=\exp(-\frac{1}{2}\int_0^T \theta_s^2 d\langle B^G \rangle_s-\int_0^T \theta_s d B^G_s),\quad  \mathcal{P}-q.s.
\end{eqnarray*}
With this representation of the density process we can apply the Girsanov theorem, stated in Appendix B.3.
Set $\phi_t=\frac{\rho_t}{\gamma_t}$ and  consider  the process 
\begin{eqnarray*}
 B^{\phi}_t= B^G_t-\int_0^t \phi_s ds, \quad t\in [0,T].
\end{eqnarray*}
By  the  Girsanov formula for $G$-Brownian motion, stated in  Appendix B.3, we deduce that $ B^{\phi} $ is
a $G$-Brownian motion under the sublinear expectation  $\mathcal{E}^\phi(\cdot)=E_G[\phi\cdot]$ and $S$ satisfies
\begin{eqnarray*}
 S_t =S_0+\int_0^t V_s d  B^{\phi}_s+ \int_0^t ( \mu_s + V_s \phi_s) d\langle B^\phi\rangle_s\quad t\in[0,T]
\end{eqnarray*}
on $(\Omega,\mathcal{H},\mathcal{E}^\phi)$. 
Since $V$ is a bounded process, the stochastic integral is a symmetric martingale under $\mathcal{E}^\phi$. 
$S$ is a symmetric $\mathcal{E}^\phi$-martingale if and only if $\mu_t + V_t \phi_t=0$.
We have shown that $\rho$ is simultaneous Radon-Nikodym type density of the EsMM-set $\mathcal{Q}=\rho \mathcal{P}$. Hence, the power set of EsMM-sets is not only the empty set since 
$\phi=\theta $. $\hfill\blacksquare$
\end{tpo}

\section{Appendix: Required results} 
In this Appendix we introduce the mathematical framework more carefully. We also collect all the results  applied in Sections 2 and 3.
\subsection{The sub order dual}
In this subsection we discuss the mathematical preliminaries for the price space of sublinear functionals for Section 3.
\newline
\textit{The topological dual space:} 
\begin{enumerate}
 \item Let  $c_{2,\mathcal{P}}$ be a capacity norm, defined in Section 2.2, on a complete separable metric space $\Omega$. Every continuous linear form $l$ on
$L^2(\mathcal{P})$ admits a representation:
\begin{eqnarray*}
 l(X)=\int X d\mu\quad \forall X\in L^2(\mathcal{P}),
\end{eqnarray*}
where $\mu$ is a  bounded signed measure defined on a $\sigma$-algebra
containing the Borel $\sigma$-algebra of $\Omega$.
If $l$ is a non-negative linear form, the measure $\mu$ is non-negative finite.
\item We have $L^2(\mathcal{P})^*=\left\{\mu=Z P: P\in\mathcal{P} \:and\:Z\in L^2(\mathcal{P})_+  \right\}$. 
\end{enumerate}
Note that the capacity norm defined in (1) is a Prohorov capacity. We apply  Proposition 3 from  \cite{bnk10}. The second assertion can be proven via a modification of Theorem I.30 in \cite{ker08}, where the case of $L^1(\mathcal{P})$ is treated.

\subsubsection{Semi lattices and their intrinsic structure} 
We begin with the most simple operation of consolidation, ignoring a subset of priors and giving a weight to the others.
\newline
\textbf{Integration:}
\newline
Let  $\mu\in \mathcal{M}_{\leq 1}(\mathcal{P})$ be the  positive measure  $\mu$ such that $\mu(\mathcal{P})\leq 1$.
 In our case the underlying space is $\Pi_{P\in \mathcal{P}}L^2(\mathcal{P})^*$ such that the density component is invariant, when considering the representation $l(X)=\int ZXdP$. So let $N\subset\mathcal{B}(\mathcal{P})$ be a Borel measurable set, a by a measure $\mu\in \mathcal{M}_{\leq 1}(\mathcal{P})$  is given by 
\begin{eqnarray*}
 \Gamma(\mu,N):\times_{P\in \mathcal{P}}L^2(\mathcal{P})^*\rightarrow  L^2(\mathcal{P})^*,\: \{\pi_P\}_{P\in\mathcal{P}}\mapsto\int_{N} 1 d\mu(P)\cdot Z.
\end{eqnarray*}
The size of $N$ determines the degree of ignorance, related to the exclusion of the prior in the countable reduction. A measure with mass less than implies an ignorance
The Dirac measure is a  projection to one certain probability model.
\\[1em]
Next, we consider the supremum operation of functionals. Note that this gives us the connection to sublinear expectations.
 \newline
 \textbf{The point-wise supremum:}
 \newline
The operation of point-wise maximum preserves the convexity.
We review a result which gives an iterated application of the Hahn-Banach Theorem.
\newline
\textit{ Representation of sublinear functionals }\cite{fri00}:
Let $\psi$ be a sublinear functional on a topological vector space $V$, then
 \begin{eqnarray*}
\psi(X)=\max_{x^*\in P_\psi} x^*(X),
\end{eqnarray*}
where $P_\psi=\{ x^*\in X^*: x^*(X)\leq \psi(X) \textit{ for all } X \in V \}\not= \emptyset$
\\[1em]
The maximum operation can also be associated to a lattice structure. In economic terms this is related to a normative choice  of the super hedging intensity.  The diversification valuation operator consolidation is set to \textit{one nonlinear} 
valuation functional. Note that the operation preserves monotonicity.

\subsection{The set of probability models}
The  model of multiple priors motivates the introduction of the following mapping
\begin{eqnarray*}
 \mathtt{c}:\mathcal{B}(\Omega)\rightarrow [0,1], \quad  \mathtt{c}(A)=\sup_{P\in\mathcal{P}}P(A).
\end{eqnarray*}
It is easy to prove that $ \mathtt{c}(\cdot)$ is  a Choquet capacity.\footnote{For a general treatment, see again \cite{dhp11}
 and the references therein.} The capacity notion may be used for an alternative formulation of Theorem 2.
\newline
Fix $(\Omega,\mathcal{B}(\Omega))=(C_0([0,T],\mathcal{B}(C_0([0,T]))$. We refer to \cite{bnk10} where the state space consists of a  cadlaq path. 
We give a criterion for the weak  compactness of $ \mathcal{P}$.
Let $\sigma^1,\sigma^2:[0,T]\rightarrow \mathbb{R}$ be two measures with a Holder continuous distribution function $t\mapsto \sigma^i([0,t])=\sigma^i(t)$.
\newline
A probability measure $P$ on $\Omega$ is a martingale probability measure if the 
coordinate process is a martingale with regard to the canonical
 (raw) filtration.
\newline
Let $\sigma^1,\sigma^2$ be two measures with a Holder continuous distribution function $t\mapsto \sigma^i([0,t])=\sigma(t)$. 
\\[1em]
\textit{Criterion for weak compactness of priors}, \cite{deke07}:
  Let $ \mathcal{P}(\sigma^1,\sigma^2)$ be the set of  martingale probability measures with
\begin{eqnarray*}
d \sigma^1(t)\leq d\langle B\rangle^P_t \leq  d \sigma^2(t),
\end{eqnarray*}
where $\langle B \rangle^P$ is the quadratic variation of $B$ under $P$. 
Then the set $ \mathcal{P}(\sigma^1,\sigma^2)$ is weakly compact.
\newline
Now, we discuss the concept of countable reduction. We apply the  following result  in Section 2. 
\\[1em]
\textit{Countable reduction}, \cite{bnk10}:
 Let $c_{2,\mathcal{P}}$ be given by a weakly compact set of probability measures $\mathcal{P}$. Then there is a countable set 
$(P_n)_{n\in\mathbb{N}} \subset \mathcal{P}$ such that for all $X\in L^2(\mathcal{P})$
\begin{eqnarray*}
c_{2,\mathcal{P}}(X)=\sup_{n\in \mathbb{N}} E^{P_n}[\vert X\vert^2]^{\frac{1}{2}}.
\end{eqnarray*}
The associated Banach spaces are the same.  This assertion holds, since the closure of $\mathcal{P}$ has a countable dense subset (for the weak${}^*$-topology or in probabilistic terms the vague topology).
\\[1em]
Following, we introduce an equivalence class associated with the $c_{2,\mathcal{P}}$-norm on $\mathcal{P}$. We start with some single prior considerations, taken from \cite{bnk10}. Note that $L^2(\left\{ P \right\})=L^2(P)$, let $ Q\in L^1(P)^*$ and remember
\begin{eqnarray*}
Q\sim P \textit{ if and only if }\bigg( \forall X\in L^1(P)_+, X=0 \: in\: L^1(P)  \Leftrightarrow \int X d Q \bigg).
\end{eqnarray*}
Whenever $\mathcal{P}$ is weakly relatively compact, we can  associate a probability measure $P$
 to $L(\mathcal{P})$, characterizing the (quasi sure) null elements in the positive cone $L^2(\mathcal{P})_+$.
Let  $ \mathcal{M}^+(c_{2,\mathcal{P}})$ be the set of non-negative finite
measures on $(\Omega,\mathcal{B}(\Omega))$ defining an element of $L^2(\mathcal{P})^*$.
 Define on $ \mathcal{M}^+(c_{2,\mathcal{P}})$ the relation $R_{c_{2,\mathcal{P}}}$ by: 
\begin{eqnarray*}
  \mu R_{c_{2,\mathcal{P}}} \nu \textit{ if and only if } \bigg(   \forall X\in L^2(\mathcal{P})_+,\int X d\mu =0 \Leftrightarrow \int X d\nu=0\bigg)
\end{eqnarray*}
It follows that $R_{c_{2,\mathcal{P}}}$ is an equivalence relation on $\mathcal{P}$. We are able to say more
about the dual space of $L^2(\mathcal{P})$.
\\[1em]
\textit{Reference measure for the positive cone, }\cite{bnk10}:
  There is a unique $R_{c_{2,\mathcal{P}}}$ equivalence class in $\mathcal{M}^+(c_{2,\mathcal{P}})$
 such that $\mu\in\mathcal{M}^+(c_{2,\mathcal{P}})$ belongs to this class if and only if
\begin{eqnarray*}
\forall X \in L^2(\mathcal{P})_+ , \{\mu(X) = 0\} \textit{ if and only if } \{X = 0\:\: in \:\: L^2 (\mathcal{P})\}.
\end{eqnarray*}
This class is referred as the canonical $c_{2,\mathcal{P}}$-class.
For every countable weakly relatively compact set $(P^n)_{n\in\mathbb{N}}$ such that (1) holds,
for $\alpha_n>0$, for each $n\in\mathbb{N}$, such that $\sum_{n\in\mathbb{N}} \alpha_n=1$ the probability measure $\sum_{n\in\mathbb{N}} \alpha_n P^n$ belongs  to the canonical $c_{2,\mathcal{P}} $-class. 
\newline
This gives us an easy definition of $\mathcal{P}$-arbitrage, as mentioned in Section 3.2.

%\subsection{A detour of stochastic analysis with a general volatility uncertainty}

%The concept of time consistency is a natural benchmark for plausibility of dynamic decision theory, see \cite{ri04} for an axiomatization in discrete time. %For %instance, in order to guarantee this property we  assume the following closedness property.
%The set of priors $\mathcal{P}$ is \textit{maximally chosen}:
 %$\mathcal{P}$ consists of all  $ P\in\mathcal{P}$ satisfying
%\begin{align*}
%E^P [X]\leq \sup_{P'\in \mathcal{P}}E^{P'} [X], \quad for\: all\: X\in L^2(\mathcal{P})
%\end{align*}
%Dynamic consistency in dynamic decision theory is discussed  in [DuEp 92]. With multiple priors the "minimal" prior may vary over time and state of %the world. 

%The definition below requires a martingale notion. The multiple and singular prior framework requires
 %a dynamic and nonlinear notion of martingales, which guarantee consistency in conditional updating.
%In the Appendix we reconsider ....+2BSDE
%\newline
%dynamic sublinear expectation
%\begin{align*}
%\mathcal{E}^o_s(X)={\esssup_{Q'\in\mathcal{P}(s,P)} }^P E^{Q'}[X|\mathcal{F}_s]\quad P-a.s.
%\end{align*}
%connection to the general case...weak cp with thm 4 in app a.1..time $t=0$
%\begin{align*}
 %\mathcal{E}^o_0(X)=\mathbb{E}^{\mathcal{P}}[X]
%\end{align*}
%after the cadlag $\mathcal{P}$-modification procedure [SoNu,.. ]
%\begin{align*}
%\mathcal{E}^o_s(X)=\mathcal{E}_s(X) \quad \mathcal{P}-q.s.\quad \textnormal{ for all } t\in[0,T]
%\end{align*}
%[SoNu 10] framework

%\begin{align*}
%P^\alpha=P^0\circ (X^{\alpha})^{-1}
%\end{align*}

\subsection{Stochastic analysis with $G$-Brownian motion}
We introduce the notion of sublinear expectation
for the $G$-Brownian motion. This includes the concept of $G$-expectation, the It\^o calculus with $G$-Brownian motion and
related results concerning the representation of $G$-expectation and (symmetric) $G$-martingales. For a more precise detour we refer to the Appendix of \cite{vor10}  and to references therein. 
\newline
At the end of this section we present a Girsanov theorem for $G$-Brownian motion, which we apply in Theorem 3 of Subsection 3.4.

Let $\Omega\neq\emptyset$ be a given set. Let $\mathcal{H}$ be a
linear space of real valued functions defined on $\Omega$ with
$c\in\mathcal{H}$ for all constants $c$ and $|X|\in\mathcal{H}$ if
$X\in\mathcal{H}$. Note that in our model  we choose $\mathcal{C}_b(\Omega)=\mathcal{H}$ and $\Omega=\Omega_T=C_0([0,T])$.
\newline
A sublinear expectation $\hat{E}$ on
$\mathcal{H}$ is a functional $\hat{E}:\mathcal{H}\rightarrow
\mathbb{R}$ satisfying  monotonicity, constant preserving, sub-additivity and positive homogeneity.
The  triple $(\Omega,\mathcal{H},\hat{E})$ is called a \textit{sublinear
expectation space}.
For the construction of the $G$-expectation, the notion of independence and $G$-normal distributions we refer to \cite{pe10}.
\newline
A process $(B_t)_{t\geq 0}$ on a sublinear expectation space
$(\Omega,\mathcal{H},\hat{E})$ is called a $G$--\textit{Brownian motion} if
the following properties are satisfied:
\begin{itemize}
\item [(i)] $B_0=0$.
 \item [(ii)] For each $t,s\geq 0$: $B_{t+s}-B_t\sim B_t$ and
$\hat{E}[\vert B_t\vert^3]\rightarrow 0$ as $t\rightarrow 0$.
 \item [(iii)]
The increment $B_{t+s}-B_t$ is independent from
$(B_{t_1},B_{t_2},\cdots,B_{t_n})$ for each $n\in\mathbb{N}$ and
$0\leq t_1\leq\cdots\leq t_n\leq t$. 
\item [(iv)]
$\hat{E}[B_t]=-\hat{E}[-B_t]=0\quad\forall t\geq 0$.
\end{itemize}
The following observation is important for the characterization of
$G$--martingales. 
 The space $C_{l,Lip}(\mathbb{R}^n)$, where $n\geq 1$ is the space of all real-valued
continuous functions $\varphi$ defined on $\mathbb{R}^n$ such that
$|\varphi(x)-\varphi(y)|\leq C(1+|x|^k+|y|^k)|x-y|~~\forall
x,y\in\mathbb{R}^n$. We define
\begin{align*}
L_{ip}(\Omega_T):=\{\varphi(B_{t_1},\cdots,B_{t_n})|n\in\mathbb{N},t_1,\cdots,t_n\in
[0,T],\varphi\in C_{l,Lip}(\mathbb{R}^n)\}.
\end{align*}
The It\^o integral can also be defined for the
following processes: Let $H_G^0(0,T)$ be the
collection of processes $\eta$ having the following form: For a
partition $\{t_0,t_1,\cdots,t_N\}$ of $[0,T], N\in\mathbb{N}$, and
$\xi_i\in L_{ip}(\Omega_{t_i})~ \forall i=0,1,\cdots,N-1$, let
$\eta$ ( see \cite{so09}) be given by
\begin{align*}
\eta_t(\omega):=\sum_{j=0}^{N-1}\xi_j(\omega)
1_{[t_j,t_{j+1})}(t)\quad\forall t\leq T.
\end{align*}
For  $\eta\in H_G^0(0,T)$ let $\Vert\eta\Vert_{M^2_G}:=
\left(E_G\left[\int_0^T|\eta_s|^2ds\right]\right)^{\frac{1}{2}}$
and denote by $M_G^2(0,T)$ the completion of $H_G^0(0,T)$ under
this norm.
\newline
 As before we can construct
It\^o's integral $I$ on $H_G^0(0,T)$ and extend it to $M_G^2(0,T)$
continuously, hence $I: M_G^2(0,T)\rightarrow
L^2(\mathcal{P})$.
\newline
The next result is an It\^o formula. The presentation of  basic notions on stochastic calculus with respect to $G$-Brownian motion lies beyong the scope of this appendix.
\\[1em]
\textit{It\^o-formula}, \cite{lipe11}:
Let $\Phi\in C^2(\mathbb{R})$ and $d X_t = \mu_t d\langle B^G \rangle_t + V_t d B^G_T,\quad t\in [0,T]$,  $\mu,V\in M^2_G(0,T)$
are bounded processes. Then we have for every $t\geq0$:
\begin{align*}
\Phi(X_t)-\Phi(X_s)=\int_s^t \partial\Phi(X_u)V_u dB^G_u + \frac{1}{2}\int_s^t  \partial\Phi(X_u)\mu_u+  
\partial^2{\Phi}(X_u)V^2_u  d\langle B^G\rangle_u.
\end{align*}
Next, we introduce a in $G$-framework of martingales.
A process $M=\{M_t\}_{t\in [0,T]}$ with values in $L^2(\mathcal{P})$ is called
$G$-martingale if $E_G(M_t|\mathcal{F}_s)=M_s$ for all $s,t$ with
$s\leq t\leq T$. If $M$ and $-M$ are both $G$--martingales $M$ is
called a symmetric $G$--martingale. This terminology also applies to general sublinear expectations  as those in Section 3.2.
\newline
By means of the characterization of the conditional $G$-expectation
we have that $M$ is a $G$-martingale if and only if for all $0\leq
s\leq t\leq T, P\in\mathcal{P}$,
\begin{align*}
M_s=\esssup_{Q'\in\mathcal{P}(s,P)}E^{Q'}[M_t|\mathcal{F}_s]\quad
P-a.s.
\end{align*}
In \cite{so09}, this identity declares that a $G$-martingale $M$
can be seen as a multiple prior martingale which is a
supermartingale for any $P\in\mathcal{P}$ and a martingale for an
optimal measure.% from the set $\mathcal{P}$.
\\[1em]
\textit{ Characterization for G-martingales}, \cite{stz11}:
Let $x\in\mathbb{R},z\in M^2_G(0,T)$ and $\eta\in M_G^1(0,T)$.
Then the process 
\begin{align*} M_t:= x+\int_0^tz_sdB_s+\int_0^t
\eta_sd\langle B\rangle_s-\int_0^t 2G(\eta_s)ds, \quad t\leq T,
\end{align*}
is a $G$--martingale.
\newline
In particular, the nonsymmetric part
${-K_t:=\int_0^t\eta_sd\langle B\rangle_s-\int_0^t2G(\eta_s)ds,}$
$t\in [0,T],$ is a $G$-martingale which is different
compared to classical probability theory since $\{-K_t\}_{t\in[0,T]}$ is
continuous, and non-increasing with a quadratic variation equal to
zero.
$M$ is a symmetric $G$--martingale if and only if $K\equiv 0$.
\\[1em]
\textit{Martingale representation}, \cite{so09}:
Let $\xi\in L_G^2 (\Omega_T)$. Then the
$G$--martingale $X$ with $X_t:=E_G[\xi|\mathcal{F}_t],t\in [0,T]$,
has the following unique representation
\begin{align*}
X_t=X_0+\int_0^tz_sdB_s-K_t
\end{align*}
where $K$ is a continuous, increasing process with $K_0=0, K_T\in
L_G^\alpha(\Omega_T), z\in H_G^\alpha(0,T), \forall \alpha\in
[1,2)$, and $-K$ a $G$--martingale. Here, $H_G^\alpha(0,T)$ is the completion of $H_G^0(0,T)$ under the norm
$\Vert\eta\Vert_{H^\alpha_G}:=
\left(E_G\left[\int_0^T\vert\eta_s\vert^2ds\right ]^{\frac{\alpha}{2}}\right)^{\frac{1}{\alpha}}$.
\newline
If is  $\xi$ bounded from above we get that $z\in
M^2_G(0,T)$ and $K_T\in L^2_G(\Omega_T)$, see  \cite{so09}.
\newline
Finally we establish a Girsanov type theorem with $G$-Brownian motion.
Consequently we establish the result, and we discuss some heuristics in terms of a $G$-Doleans Dade exponential.
Define the density process by $\mathtt{E}^\theta$ as the unique solution of $d \mathtt{E}^{\theta}_t=\mathtt{E}^{\theta}_t \theta_t dB^G_t$,  $ \mathtt{E}^{\theta}_0=1$.
The proof of the  Girsanov theorem is based on a  Levy martingale characterization theorem for $G$-Brownian motion.
\\[1em]
\textit{Girsanov for G-expectation}, \cite{xsz11}:
 Assume the following Novikov type condition:
There is an $\epsilon>0$ such that
\begin{align*}
 E_G\left[\exp((\frac{1}{2}+\epsilon)\int_0^T \theta^2_s d\langle B^G\rangle_s)\right]< \infty
\end{align*}
 Then $B^{\theta}_t= B^G_t-\int_0^t \theta_s \langle B^G\rangle_s$   is a $G$-Brownian motion under
 the sublinear expectation $\mathcal{E}^\theta(\cdot)$ given by, 
$ \mathcal{E}^\theta(X)=E_G[\mathtt{E}^{\theta}_T \cdot X]$, $\mathcal{P}^{\theta}= \mathtt{E}^{\theta}_T\mathcal{P}$ with $X\in L^2(\mathcal{P}^{\theta})$.

\end{appendix}

\bibliographystyle{econometrica}

\ifx\undefined\BySame
\newcommand{\BySame}{\leavevmode\rule[.5ex]{3em}{.5pt}\ }
\fi
\ifx\undefined\textsc
\newcommand{\textsc}[1]{{\sc #1}}
\newcommand{\emph}[1]{{\em #1\/}}
\let\tmpsmall\small
\renewcommand{\small}{\tmpsmall\sc}
\fi

 %\bibliography{bib}

\end{document}